\documentclass[acmsmall]{acmart}

\AtBeginDocument{%
  }

\setcopyright{acmlicensed}
\acmJournal{PACMHCI}
\acmYear{2025} \acmVolume{9} \acmNumber{2} \acmArticle{CSCW051} \acmMonth{4}\acmDOI{10.1145/3710949}

\begin{document}

\title{Connecting through Comics}
\subtitle{Design and Evaluation of Cube, an Arts-Based Digital Platform for Trauma-Impacted Youth}


\author{Ila K Kumar}
\authornote{Both authors contributed equally to this manuscript.}
\email{ilak@media.mit.edu}
\orcid{0000-0002-3482-6615}
\author{Jocelyn Shen}
\authornotemark[1]
\email{joceshen@media.mit.edu}
\orcid{0000-0001-7809-5474}
\affiliation{%
  \institution{Massachusetts Institute of Technology}
  \streetaddress{75 Amherst St}
  \city{Cambridge}
  \state{MA}
  \country{USA}
  \postcode{02139}
}

\author{Craig Ferguson}
\email{fergusoc@media.mit.edu}
\affiliation{%
  \institution{Massachusetts Institute of Technology}
  \streetaddress{75 Amherst St}
  \city{Cambridge}
  \state{MA}
  \country{USA}
  \postcode{02139}
}

\author{Rosalind W Picard}
\email{picard@media.mit.edu}
\affiliation{%
  \institution{Massachusetts Institute of Technology}
  \streetaddress{75 Amherst St}
  \city{Cambridge}
  \state{MA}
  \country{USA}
  \postcode{02139}
}

\renewcommand{\shortauthors}{Ila K Kumar, Jocelyn Shen, Craig Ferguson, and Rosalind W. Picard}

\begin{abstract}
This paper explores the design, development and evaluation of a digital platform that aims to assist young people who have experienced trauma in understanding and expressing their emotions and fostering social connections. Integrating principles from expressive arts and narrative-based therapies, we collaborate with lived experts to iteratively design a novel, user-centered digital tool for young people to create and share comics that represent their experiences. Specifically, we conduct a series of nine workshops with $N=54$ trauma-impacted youth and young adults to test and refine our tool, beginning with three workshops using low-fidelity prototypes, followed by six workshops with Cube, a web version of the tool. A qualitative analysis of workshop feedback and empathic relations analysis of artifacts provides valuable insights into the usability and potential impact of the tool, as well as the specific needs of young people who have experienced trauma. Our findings suggest that the integration of expressive and narrative therapy principles into Cube can offer a unique avenue for trauma-impacted young people to process their experiences, more easily communicate their emotions, and connect with supportive communities. We end by presenting implications for the design of social technologies that aim to support the emotional well-being and social integration of youth and young adults who have faced trauma.

\end{abstract}



\received{January 2024}
\received[revised]{July 2024}
\received[accepted]{October 2024}

\begin{teaserfigure}
  \includegraphics[width=\textwidth]{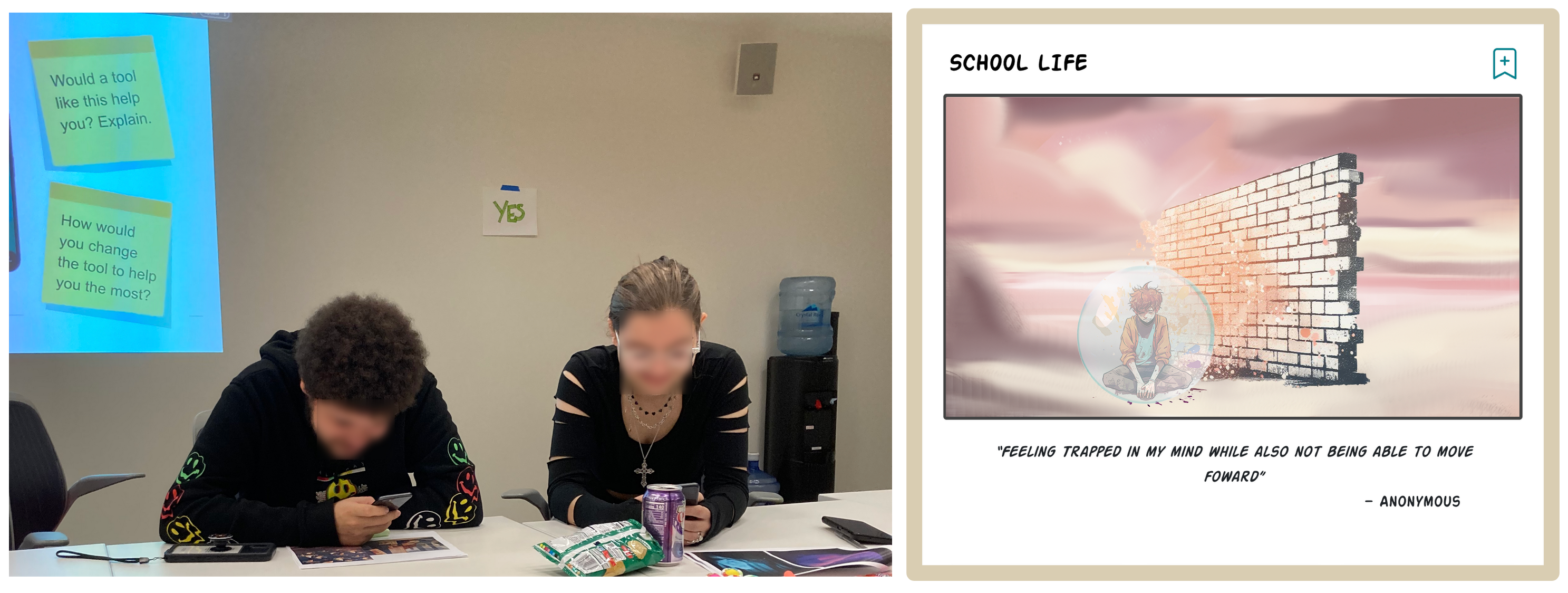}
  \caption{We ran a series of workshops with $N=54$ trauma-impacted youth and young adults to design a digital platform that supports emotion processing, self-expression, and social connection through expressive, arts-based therapies.}
  \Description{On the left is a photo of two participants during the workshop interacting with our prototype on their smartphones. On the right is an example comic that a participant created. It shows a boy sitting cross-legged in the sand, with a burning wall right behind him. He is trapped inside a bubble. The comic description reads ``feeling trapped in my mind while also not being able to move forward.''}
  \label{fig:teaser}
\end{teaserfigure}

\maketitle
\section{Introduction}
New solutions are needed to support the estimated 20-48\% of youth in the United States who have been exposed to multiple types of trauma in their childhood \cite{saunders_epidemiology_2014}. These numbers may be under-estimates, as many studies leave out non-violent traumatic events (often interpersonal), such as neglect, psychological maltreatment, attachment separations and impaired caregiving systems \cite{van_der_kolk_body_2014, blaustein_treating_2010}. Childhood adversity increases risk for many problems downstream, including mental illness, chronic illness, incarceration, and unemployment \cite{felitti_relation_2002}. For youth who experience repeated instances of trauma, there are significant long-term consequences on their development, which can include deficits in emotional development, causing youth to have difficulty understanding what they feel, where these feelings come from, how to cope with them, and how to express them \cite{blaustein_treating_2010}. Because many of the traumas they experience are interpersonal and involve their earliest connections, youth also struggle with trusting others and developing social connections \cite{blaustein_treating_2010, herman_trauma_2015}. Trauma early in life can also cause delays in cognitive development, impacting youths’ ability to express themselves through language and making it harder for them to express how they feel to others and build relationships \cite{blaustein_treating_2010}. This is particularly problematic given that social support has been shown to directly reduce the severity of psychological symptoms, and protect individuals from the potentially adverse effects of stressful situations \cite{cohen_stress_1985,herman_trauma_2015,kawachi_social_2001}. 

To support young people in developing foundational emotional skills and protective social connections, experts advocate for helping them learn to identify and express their emotions in a safe environment \cite{blaustein_treating_2010}. This includes helping youth become aware of and understand their different internal states (affect identification) and helping them build the skills and confidence to be able to effectively and safely share emotional experiences with others \cite{blaustein_treating_2010}. This corrects the maladaptive belief that sharing emotional experiences is dangerous or unhelpful, and helps encourage them to build safe relationships throughout their life \cite{blaustein_treating_2010}. Scholars also emphasize the importance of having trauma survivors hear about the emotional experiences of others, so they feel a sense of commonality with a larger community, which can make them more open to social connections \cite{herman_trauma_2015}. 

Prior works in human-computer interaction (HCI) demonstrate that technological solutions can provide new modalities for emotion understanding and expression \cite{rajcic_mirror_2020, peng_trip_2018}. Such platforms could offer unique benefits to individuals who have experienced trauma, including anonymity and connection to support networks, especially if they utilize therapeutic and psychological precedents for helping users authentically express their emotions, for example through creative avenues \cite{malchiodi_expressive_2005}. 

To this end, we develop and evaluate Cube, a novel tool to support young people who have experienced trauma in understanding and expressing their emotions and connecting with others, utilizing expressive and narrative-based therapy techniques. We begin by qualitatively analyzing feedback from three workshops in which trauma-impacted youth and young adults interacted with low-fidelity prototypes of this tool. We then present a web version of the tool (Cube), designed based on the feedback received on the low-fidelity prototypes. Finally, we qualitatively and quantitatively analyze feedback from six workshops in which participants interacted with Cube and conduct an empathic relations analysis on the workshop artifacts to understand the potential impact of the system on users' sense of connectedness when used at scale.

In summary, we aim to answer the following research question:
\textbf{How does Cube, a platform for sharing emotional experiences via comic creation, impact emotion expression and a feeling of connection among youth who have experienced trauma?} Our contributions include:
\begin{itemize}
    \item An iterative design process towards the creation of a digital platform that supports trauma-impacted youth in (1) emotion processing and expression and (2) fostering sense of connection, inspired by expressive and narrative therapy approaches.
    \item Platform specifications and implementation of Cube, a novel digital tool utilizing comic creation to facilitate the expression of and engagement with emotional narratives.
    \item Qualitative and quantitative results from the youth's interactions during workshops with the prototypes and the resulting app.
    \item Design guidelines for future emotional expression and social support platforms geared towards trauma-impacted youth, motivated by the youth's feedback.
\end{itemize}

\section{Related Works}

\subsection{Expressive Therapies}

Expressive therapies use creative arts practices to help individuals process difficult experiences and express complex emotions. These techniques can be particularly helpful for young people who have experienced trauma, as they focus on non-verbal expression and externalizing emotional experiences \cite{malchiodi_expressive_2005}.

By asking them to create a visual representation of what’s on their mind, \textbf{art therapy} has been shown to help youth externalize thoughts, feelings, and experiences \cite{bosgraaf_art_2020}. This can shift troubling experiences from the present to the past and make it easier for young people to talk about these experiences. Because art therapy involves action-oriented activities (creating things), it can tap into the limbic system’s sensory memory of traumatic events and help individuals bridge their implicit and explicit memories \cite{malchiodi_creative_2014}. In addition, art therapy activities can be calming and relaxing, which can help people enter a more reflective state \cite{malchiodi_creative_2014}.

\textbf{Sandtray therapy}, another expressive therapy, helps young people express emotional issues that are difficult to verbalize \cite{malchiodi_expressive_2005, sweeney_sandtray_2021}. In this type of therapy, clients are asked to represent their experiences by creating scenes with provided miniature objects and figures. This projection (putting emotions onto the scene) and symbolization (representing people and challenges with objects) can help with externalization of experiences, while reducing pressure for those who lack confidence in their creative or artistic abilities \cite{malchiodi_expressive_2005}.

When expressive art therapies are done with groups of young people, it can help participants feel more connected to others and less alone in their experiences. Sharing their emotions and creations with others in the session helps individuals build emotional expression skills and gain confidence by doing so in a safe space. Members also see the ways in which they are similar to others through the focus on universal emotional experiences. By encouraging others to respond positively, such activities can also increase social support for those in the group \cite{malchiodi_expressive_2005, yalom_theory_1995}. In particular, prior works indicate a need for digital tools that support art therapy through supplementing the creation and sharing process \cite{cornejo_vulnerability_2016, mihailidis_towards_2010}. Our work on the development of Cube, a digital platform for personal narrative construction, draws on art therapy approaches to empower trauma-impacted youth to express their stories. 

\subsection{Narrative-Based Therapies}

Beyond visual arts-based approaches, stories hold great power for helping a person assess how they feel, think, and behave in everyday life, which makes personal storytelling an effective tool in therapy \cite{blenkiron_stories_2005, mateas_narrative_nodate}. In particular, storytelling has been used in cognitive behavioral therapy (CBT), a conversational therapy that aims to change the way a person feels by recognizing unhelpful thinking and how this affects their mood \cite{beck_cognitive_1976}. Telling a personal story might allow an individual to externalize their emotions and examine them more objectively or think through a change process \cite{otto_stories_2000, torre_putting_2018}. The use of stories and metaphors in CBT has been shown to improve clinical outcomes through increased memorability of issues in one’s thinking \cite{blenkiron_stories_2005,martin_therapists_1992}. Reconstructing pictorial narratives through computational means has been shown to support emotional self-awareness for middle school girls \cite{daily_girls_nodate}.

\textbf{Narrative therapy} is a type of therapy that encourages patients to identify and rewrite problem-saturated narratives, similar to cognitive reframing \cite{white_narrative_1990,hayward_what_nodate,nurser_personal_2018}. 
Narrative therapy is effective in reducing symptoms of depression, with results comparable to cognitive-behavioral therapy \cite{lopes_narrative_2014}. Notably, an important piece of narrative therapy is the way in which a patient tells their life story. Methods that draw on storytelling and word play are employed to elicit stories from patients \cite{ricks_my_2014, yoosefi_looyeh_treating_2014, mcardle_fiction_2001}. For example, \citet{yoosefi_looyeh_treating_2014} use metaphors and fairy tales to describe and initiate discussion about social situations in order to help treat social phobia in children. Other works use movies, photographs, or song lyrics to discuss and analyze characters, including their emotions, feelings about self and others, and experiences, and discuss how clients can relate \cite{ricks_my_2014}. Patients may use this experience as a beginning in telling their story and how their story could be remade or reshaped.

\textbf{Bibliotherapy} is a method of therapy that uses interaction with stories to identify with characters in narratives and vicariously experiencing their emotions \cite{lenkowsky_bibliotherapy_1987, marrs_meta-analysis_1995, madryas_narrative_nodate}. This use of stories can help young people distance themselves from their emotions by talking about characters and finding a focus outside themselves
\cite{malchiodi_creative_2014}. Bibliotherapy has been shown to be effective for problems like communication skills and assertiveness in particular \cite{smith_use_1987}, making it a valuable method for youth and young adults who have experienced trauma.

\subsection{Digital Platforms for Connection and Emotion Expression}
Today, many personal stories are shared through social media, which can also serve as a mode of self expression. However, social media usage can have a damaging effect on the sense of connection and well-being of young people who have experienced trauma. Prior work indicates that online risks such as explicit photo sharing and cyberbullying evoke symptoms of PTSD \cite{mchugh_when_2018}. Furthermore, social networks do not scaffold the process of sharing emotions or responding appropriately to emotions, and the results can be detrimental to individual mental health. Prior work shows that unstructured participation in online mental health forums can actually lead to greater stress \cite{kaplan_internet_2011}. While some social platforms have been designed to scaffold peer-to-peer support \cite{morris_efficacy_2015}, few have supported safe emotional self-expression through drawing on expressive and narrative-based therapies.

Several HCI efforts have presented digital interfaces for aiding self-awareness and emotion processing, independent of social components. Early work used simple scales for mood reporting \cite{morris_mobile_2010}, and more recently, creative ways to present user information in a way that stimulate self reflection. These interfaces often use elements of narratives to structure information delivery in a way that is natural and engaging to the user, and have been shown to lead to tangible behavior change in activities that improve mental well-being \cite{saksono_storymap_2021}. Prior work has also used expressive writing, reconstruction of emotions, and visualization of physiological data to raise self awareness of one's physical and mental state\cite{daily_inner-active_2004}.  Other example interfaces include intelligent mirrors that generate emotionally-relevant poetry \cite{rajcic_mirror_2020}, personalized animated movies for self reflection \cite{peng_trip_2018}, and visual user interfaces as emotional prosthetics to improve emotional memory \cite{mcduff_affectaura_2012}. 

While these works demonstrate the efficacy of computational tools to scaffold emotional reflection, few studies explore how to generate outputs that are meaningful for others to interact with in a social context. One such work conducted design studies to understand how stories shared through text message can improve mental health outcomes and feelings of general connection to others \cite{bhattacharjee_i_2022-1}. Other works within the CSCW community explore how digital platforms can be used to scaffold ``collective narratives'' for groups of people with shared experiences \cite{cong_collective_2021}. To our knowledge, our work extends existing literature by exploring this topic within the context of trauma-impacted youth, a critical intersection given that engagement with digital platforms for trauma-impacted youth appears to come with unique considerations \cite{badillo-urquiola_social_2020, badillo-urquiola_understanding_2017}. 
% We begin to fill this gap by designing and evaluating a digital platform that uses creative and therapeutic techniques to support young people who have experienced trauma in processing and sharing their emotions with similar peers.  

In particular, our work focuses on a population of youth who have experienced developmental trauma and how to design digital interventions to support (1) emotion processing and expression and (2) sense of connection. Digital storytelling has the potential to help trauma-impacted youth with emotion processing \cite{anderson_collaborative_2012}, since traumatic memories are less organized than other memories \cite{foa_cognitive_2004} and may need more scaffolding afforded by technological solutions \cite{oleary_design_2017, zhang_separate_2022}. Additionally, marginalized communities often naturally turn to online communities to seek support for challenges that are under-discussed in traditional settings due to existing social infrastructure \cite{devito_social_2019, andalibi_lgbtq_2022}. Social platforms have the potential to offer anonymity, which can encourage more authentic self disclosure and engagement with other people's experiences, garnering more social support
\cite{yang_channel_2019, andalibi_social_2018, choudhury_mental_2014}. Stories shared within specific communities can also enable users to seek and provide social support in targeted and scalable ways \cite{progga_just_2023}.
For trauma-impacted youth, the anonymity, scaffolding, and broad reach offered by digital platforms could support safe emotional disclosure as well as connection to the experiences of other similar youth. Our work is, to the best of our knowledge, the first work that focuses on how interdisciplinary methods in expressive and narrative therapies combined with design of digital social platforms can support the marginalized community of trauma-impacted youth. 

% This corrects the maladaptive belief that sharing emotional experiences is dangerous or unhelpful, and helps encourage them to build safe relationships throughout their life \cite{blaustein_treating_2010}. Scholars also emphasize the importance of having trauma survivors hear about the emotional experiences of others, so they feel a sense of commonality and like they are part of a larger community, which can make them more open to social connections \cite{herman_trauma_2015}. ”

% In our iterative design process, we elucidate how participant feedback shifts when using traditional paper prototypes compared to a digital app for creative expression of personal experiences, highlighting the need for digital tools in this domain. 

% Compared to prior CSCW works on designing social platforms broadly, we focus on how interdisciplinary methods in expressive and narrative therapies as well best practices in the design of digital social platforms can support the marginalized community of trauma-impacted youth. 

% (1) to the best of our knowledge, no prior works use Design is informed through trauma-impacted lens and therapeutic practices that specifically help traumatized youth
% (2) explore how design affordances from digital tools can actually support this population

% While these works demonstrate the efficacy of computational tools to scaffold emotion expression, they are not sufficient for social platforms that are designed for youth who have experienced developmental trauma. 

\section{Methods}
We begin with an iterative design approach with lived experts -- trauma-impacted youth and young adults -- to develop and refine our tool. In the following section, we discuss our participant recruitment, study procedure, and data collection and analysis methods. 

\subsection{Participants}

Fifty-four young people participated in the research study, with 25 interacting with a low-fidelity prototype (15 virtually and 10 in-person) and 34 interacting with the Cube web app (16 in-person and 18 virtually). Five participants interacted with both the low-fidelity prototype and Cube web app. Participants were between the ages of 14 and 28, with 24\% ($N = 13$) aged 14 - 17, 39\% ($N = 21$) aged 18 - 21, 28\% ($N=15$) aged 22 - 25, and 9\% ($N=5$) aged 26 - 28. Participants were located across four states: California, Georgia, Massachusetts and Rhode Island. All participants had experienced trauma in their childhood or adolescence. Within the sample, at least 50\% ($N=27$) had interacted with the child welfare system, with at least 35\% ($N=19$) currently or formerly living in foster care due to having faced abuse or neglect in their homes.

\subsection{Study Procedure}

All research procedures, materials, and data management systems were reviewed and approved by our university's Institutional Review Board prior to implementation. 

To ethically engage participants, we partnered with four local and national organizations that support youth and young adults who have experienced traumatic events. Through these organizations, we recruited youth and young adults to participate in technology design workshops. These workshops were facilitated by a member of the research team and supported by a research assistant, who took notes during discussions. The workshops were primarily educational, focusing on teaching young people about mental health and technology and empowering them to have a voice in the technology design process. Attendees who were 14 or older and proficient in English were invited to opt-in to to have their workshop artifacts analyzed for the research project. All attendees, regardless of participation in the research component, were compensated in whatever way was deemed appropriate by the overseeing organization, ranging from \$35-50 Amazon, Uber or Walmart gift cards to \$15-20 hourly compensation. While we envision our tool ultimately being used asynchronously by youth (similar to social media usage), we chose to iteratively design and evaluate our tool through these workshops in order to reach vulnerable youth and prioritize activities that made it comfortable and easy for them to share technology feedback and ideas. 

All nine workshops began with a reminder that young people could stop participating in the workshop or the research component at any time, and should only disclose as much information as they were comfortable sharing. Then, the team conducted an icebreaker activity asking participants to pick a piece of candy and share something about themselves based on the color's prompt, to help them get to know each other and the research team. After the icebreaker, the facilitator provided educational programming on the factors that contribute to mental wellness and some of the technologies that have been created thus-far to support well-being. Drawing from Design Justice principles \cite{costanza-chock_design_2020}, the facilitator emphasized that participants were qualified and necessary for designing technology that could meet their needs, because participants are experts in their experience and are capable of making decisions (the core component of design). With this context, the team then presented participants with a series of prototypes to use and give feedback on, including a physical or digital version of the comic creator prototype. 

In the first workshop, which was virtual, participants were prompted to think about a challenge they were facing and create a 4-panel comic about a character dealing with a similar problem, using a Google slides deck (see Figure \ref{fig:loficomic}). They were then asked to read each others' comics and give feedback on the activity via a follow-up Qualtrics survey. This survey asked participants to rate and explain whether the tool would help them understand their emotions, calm themselves down, feel less alone, and/or express themselves in new ways. It also asked them to rate and explain how likely they would be to use a tool like this in the future and whether they would recommend a tool like this to other youth who have gone through similar challenges. Finally, it asked them how they would change the tool to better help them. 

In the second and third workshops, both in-person, participants were similarly prompted to create a 4-panel comic, using paper materials (see Figure \ref{fig:lofipaper}). The comic format was modified to accommodate technological limitations of the in-person workshops. The facilitator then collected all comics and distributed them amongst the participants, prompting them to read the comic they received and use the provided materials to create a responding comic in which a second character helps the main character solve the problem or makes them feel cared for (see Figure \ref{fig:lofipaperreply}). This was added after the first workshop to understand how participants felt about responding to others' comics. After interacting with the prototype, the facilitator prompted participants to share their feedback via a guided discussion.

In the following six workshops (three in-person and three virtual), participants were asked to create a comic on their mobile device using the novel web app, Cube (see Figures \ref{fig:createcomic}, \ref{fig:describecomic}, and \ref{fig:communitypage}). After interacting with the app, the facilitator prompted participants to share their feedback via a guided discussion. Due to the unique structure of one partner organization, participants in one of the virtual workshops ($N=9$) were asked to give feedback via a follow-up Qualtrics survey. This survey asked participants to rate and explain how they felt about the comics they created, whether they related to comics they read, and how they felt about the comic response options. It also asked them to explain whether the tool made it easier for them to express their feelings or problems and/or helped them feel less alone or more supported, and describe how they would change the tool to better help them. These participants were also invited to make more comics on their own to inform their feedback. 

\subsection{Data Collection and Analysis}\label{sec:dataanalysis}

For in-person workshops, verbatim discussion notes and comic visuals were collected for analysis. For virtual workshops, video recordings, audio transcripts, comic visuals, comic descriptions and survey results were collected for analysis. For all comics created using the Cube app, the system collected visual information both as PNG images and also as JSON representations of all the art assets used and their relative size and placement. Using a bottom-up thematic-analysis approach and affinity diagramming techniques, two members of the research team collaboratively analyzed the qualitative data, clustering it based on related themes, then discussing and assigning labels to each theme \cite{braun_using_2006,ulrich_kj_2003}. Because the Cube app did not require participants to log in or provide any identifiable information, the data from this source was not matched to participant IDs in analysis or presentation in Section \ref{sec:cubewebapp}. 

Next, we use quantitative analysis to (1) explore whether participants can empathize with each others' comics created using the Cube app and (2) visualize how empathetically related comics are grouped based on visual designs permitted by the Cube app. The latter question allows us to better understand what types of comics are more or less related to one another, and how participants best utilized the features of Cube.

To quantitatively observe connections between participants' comics, we embedded participant's stories using the \textsc{EmpathicStoriesBART}model from \citet{shen_modeling_2023}, which is designed specifically for modeling empathetic relationships between two pieces of text. Since much of the information needed to understand the participant's comic is in the \textit{image} they constructed, we used GPT-4\footnote{\url{https://openai.com/research/gpt-4}, last accessed January 8, 2023} to generate a \textit{textual} description of the comics and append this to the participant's written caption. In particular, we fed images of participants' comics to GPT-4 via the OpenAI API, and prompted the language model with ``\textit{Please generate a textual description of the image}.'' We then concatenated the GPT-4 generated image description with the participants' written caption, and fed this into the \textbf{EmpathicStoriesBART} model to get empathically encoded comic embeddings. 
From the embeddings outputted by the \textbf{EmpathicStoriesBART} model, we used UMAP\footnote{\url{https://umap-learn.readthedocs.io/en/latest/}, last accessed January 5, 2023} to visualize the comics in 2D space and the connections between comics, allowing insights into the relationships between user comics. 

\section{Low-fidelity Prototypes}
Before implementing a deployable version of our technology, we created and evaluated low-fidelity prototypes in a series of workshops with trauma-impacted youth.
% To evaluate the app concept prior to development, w
\begin{figure}[t]
    \centering
    \includegraphics[width=.7\textwidth]{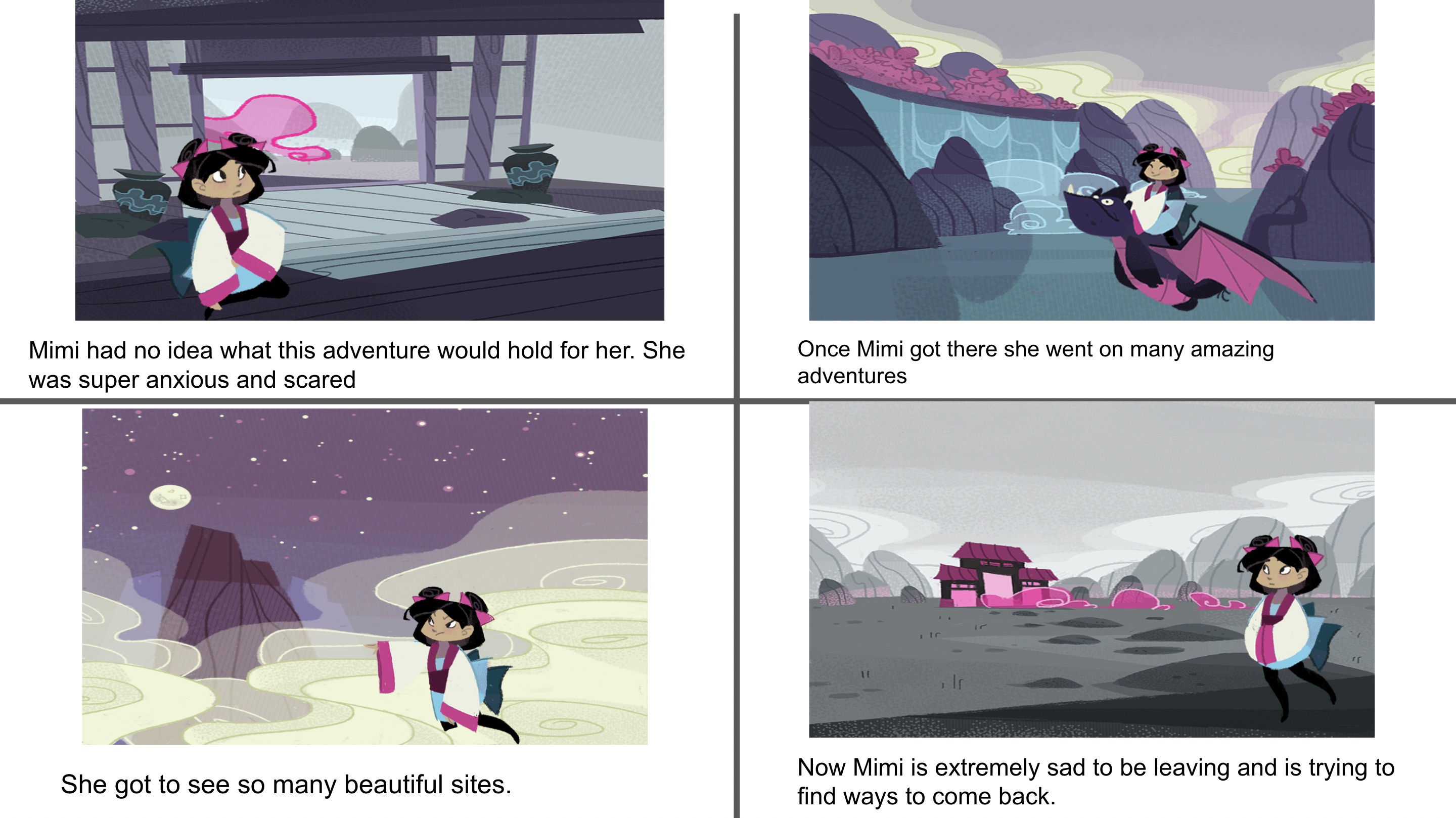}
    \caption{Example comic created using the virtual low-fidelity prototype.}
    \Description{A comic in which a girl goes on an adventure that she is apprehensive about, enjoys it, and is then sad to leave.}
    \label{fig:loficomic}
\end{figure}

\begin{figure}[t]
    \centering
    \begin{minipage}[b]{0.45\textwidth}
        \vspace{0pt}
        \includegraphics[width=\textwidth]{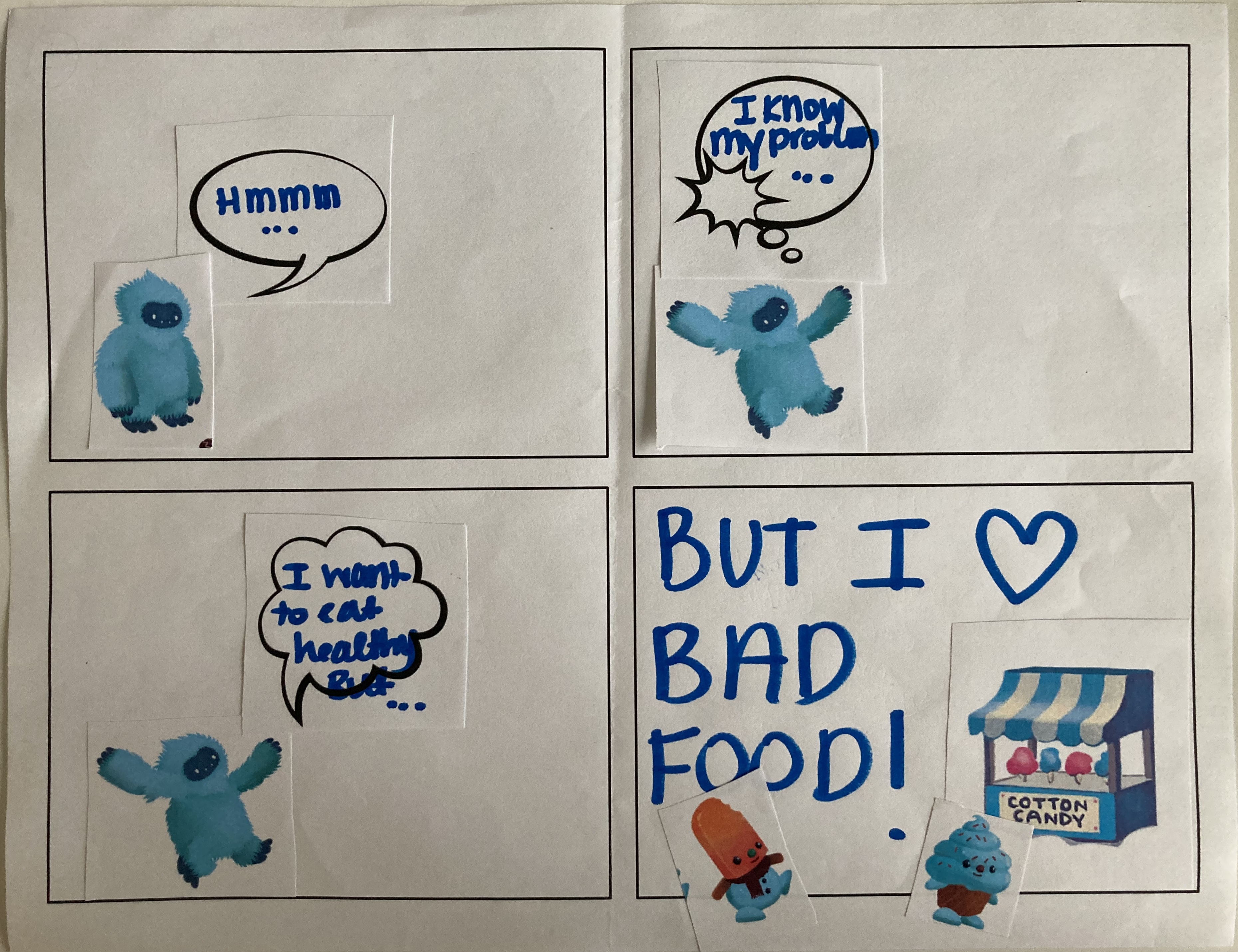}
        \caption{Example comic created using the paper low-fidelity prototype.}
        \Description{A comic in which a furry creature talks about how they want to eat healthy but love bad food, depicting the desired foods using food creatures and a cotton candy stand.}
        \label{fig:lofipaper}
    \end{minipage}
    \hfill
    \begin{minipage}[b]{0.45\textwidth}
        \includegraphics[width=\textwidth]{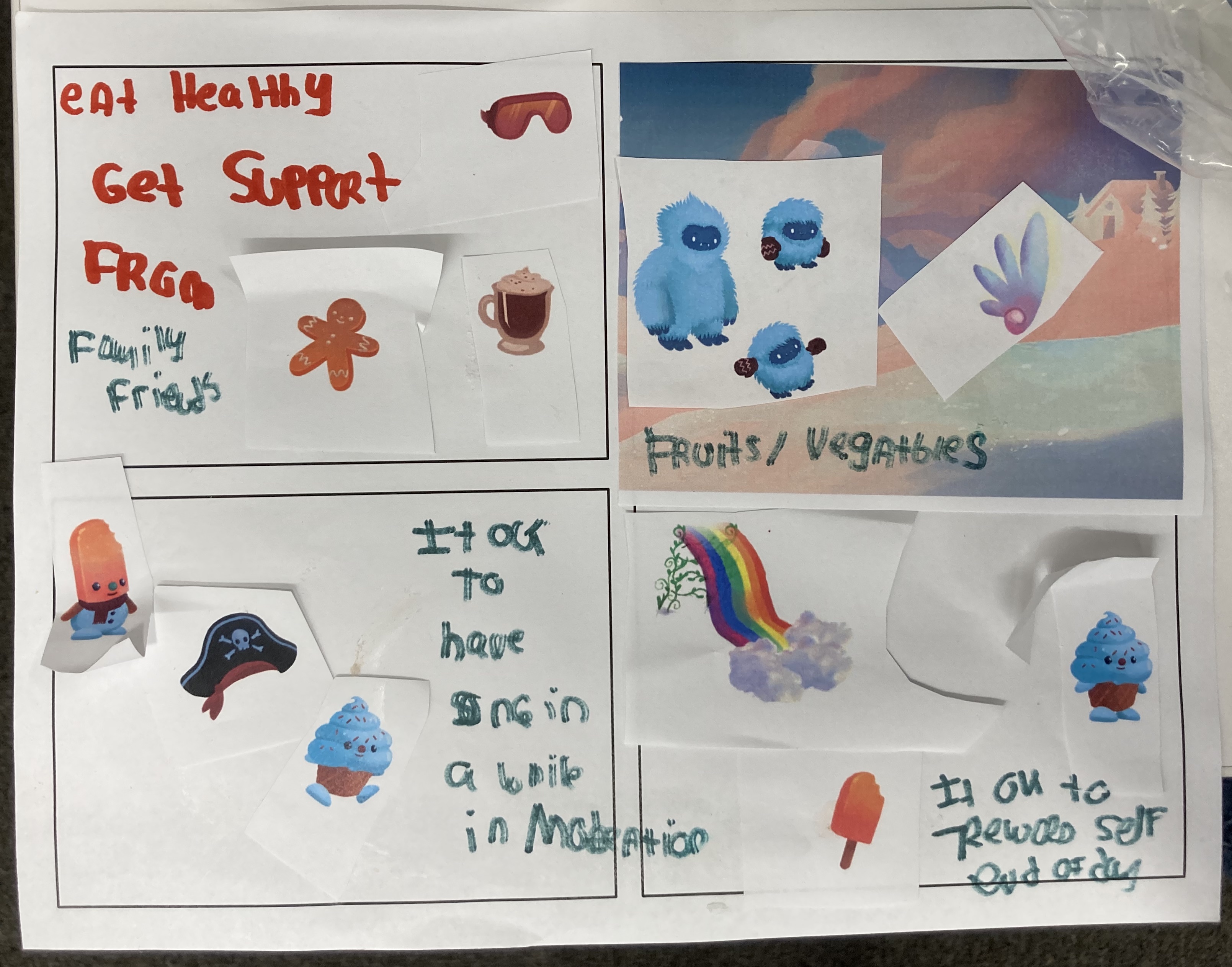}
        \caption{Example response comic created using the paper low-fidelity prototype.}
        \Description{A comic which portrays guidance and reassurance for the furry creature in Figure \ref{fig:lofipaper}.}
        \label{fig:lofipaperreply}
    \end{minipage}
\end{figure}

\subsection{Prototype Designs}

We conducted the first three workshops with low-fidelity prototypes of the app. In these prototypes, participants were asked to create a four-panel comic about a character dealing with a similar problem. For the first workshop, which was virtual, participants could create comics by dragging supplied or participant-provided visual assets into four-panel templates within a Google slides deck (for example, Figure \ref{fig:loficomic}). Supplied assets included 11 backgrounds, 20 characters, and 14 props that were created by an artist for a past storybook creation study. To accommodate the in-person nature of the second and third workshops, participants could create comics using markers, supplied visual assets, glue sticks, and 4 panel paper templates (for example, Figure \ref{fig:lofipaper} and \ref{fig:lofipaperreply}). 

Drawing inspiration from expressive therapy practices, this activity was designed to encourage self-expression and support the externalization process \cite{malchiodi_expressive_2005}. We prioritized visual expression of challenges to support with processing, expression and externalization of potentially-emotional experiences \cite{malchiodi_expressive_2005, bosgraaf_art_2020, malchiodi_creative_2014}. Additionally, we supplied a set of assets for participants to use in the visualization process to provide opportunities for projection and symbolization (to further help with externalization) and make the process less intimidating for participants who did not identify as artistic \cite{malchiodi_expressive_2005,sweeney_sandtray_2021}. Finally, in line with group art therapy techniques, the workshops included an opportunity to see, and in the second and third workshops, respond to others' comics, to help young people feel connected and validated by the commonality of their experiences \cite{malchiodi_expressive_2005, yalom_theory_1995}.

Our design was also informed by the philosophy of narrative therapy that people are separate from their traumas, and can rewrite their own stories \cite{lilly_putting_2023}. In particular, we use this design to balance creative freedom and agency over the story with scaffolding (the task of creating a comic using predefined visual assets) that reduces cognitive load and mitigates direct engagement with difficult, trauma-saturated narratives. Additionally, narrative therapists use literary tools such as metaphor to prompt discussions around mental health without inducing trauma. Following this framework, the assets we provided were intended to support participants in using symbolism and fiction to convey complex emotions without triggering overwhelming negative emotions.

\subsection{Prototype-testing Feedback}

\paragraph{Emotion Processing and Expression}

Survey feedback from the fifteen participants who interacted with the prototype virtually indicated that it was able to help them process and express their emotions. Eleven (73\%) said they somewhat ($N=6$) or strongly ($N=5$) agree that the app would help them better understand how they were feeling. Three participants further described how the app helped them think about and address their emotions. One participant (P7) explained, "\textit{It prompted me to really think deep about how I was feeling in order to find the best images to match those feelings.}" Eight participants also reported that creating a comic helped them express their feelings, saying it helped them "\textit{visually explain how [they were] feeling}" (P5) and "\textit{be more open to [their] feelings}" (P1). 

% P7 = "It prompted me to really think deep about how I was feeling in order to find the best images to match those feelings."
% P5 = "visually explain how [they were] feeling"
% P1 = "be more open to [their] feelings"

Twelve virtual prototype participants (80\%) also reported that they somewhat ($N=9$) or strongly ($N=3$) agree that the app would help them calm themselves down when they're feeling angry, sad or anxious. Two of these participants elaborated that the prototype could provide a helpful distraction from negative emotions they were experiencing, which may have contributed to their ability to express themselves. One participant (P12) explained, "drawing a comic would make me focus on a story and creativity rather than feeling sad." 
% P12 = "drawing a comic would make me focus on a story and creativity rather than feeling sad."

At the same time, nine virtual prototype participants (60\%) advocated for more supplied assets to support with emotion expression. Six of these participants generally wanted a wider variety of assets, particularly characters and backdrops. Five participants specifically wanted more characters with emotional expressions to be added. They offered up multiple ideas for how to accomplish this goal, ranging from adding more expressive characters to allowing users to customize characters with emotional facial expressions or symbols (such as hearts or tears). Importantly, participants did \textit{not} ask for the tool to include infinite visualization options, perhaps because having a limited set of assets necessarily reduced the cognitive load of an already difficult emotional processing task. As one participant (P7) explained, having a variety of supplied assets would be more helpful than having the freedom to add in their own drawings or assets; ".\textit{..having a wider selection of pre-picked images would help because sometimes its hard to think of your own to add especially when you're experiencing hard emotions.}" 
% P7 = "...having a wider selection of pre-picked images would help because sometimes its hard to think of your own to add especially when you're experiencing hard emotions."

Four participants, three who interacted with the virtual prototype (20\%) and one who interacted with the paper version (10\%), pointed out that the prototype required significant time and energy, with one (P2) explaining that "\textit{coming up with a story takes work}" and "\textit{perfectionism does not help to do this activity in a timely manner.}" These participants expressed that this commitment would decrease the frequency with which they would use the tool on their own. An additional paper prototype user said that people should be able to select the number of panels they wanted in a comic (ranging from 1 to 4), to reduce the effort required and to match the complexity of the comic to that of the challenge they were trying to express.
% P2 = "perfectionism does not help to do this activity in a timely manner"

Eight participants (7 virtual, 47\%, and 1 paper prototype user, 10\%) explicitly said that they found the prototype to be engaging and helpful, while five virtual participants (33\%) wrote that they did not think the prototype would be useful for them. Three of these participants cited that they did not like the medium (art, creative storytelling, or using digital tools for mental health), with expressions such as "\textit{Some youth find art stressful}" (P11) and "\textit{It stresses some people out having to come up with a storyline}" (P2). One participant (P13) said they were not sure if they would use the tool if they were experiencing a strong emotion, and another stated that they were not able to attach any emotions to their comic: "\textit{It helps get the creative juices going but I could not attach any emotions to it.}" 
% P11 = "Some youth find art stressful"
% P2 = "It stresses some people out having to come up with a storyline"
% P13 = "It helps get the creative juices going but I could not attach any emotions to it" 

\paragraph{Sense of Connection}

In general, participants expressed that seeing others' comics helped them feel less alone and more connected to others. Eight of the participants who interacted with the  prototype virtually (53\%) reported that they somewhat ($N=6$) or strongly ($N=2$) agree that the app would help them feel less alone in what they were going through. Two of these participants elaborated that reading others' comics helped them feel less alone. Additionally, amongst participants who interacted with the paper prototype and were asked to respond to another participants' comic, four (40\%) talked about how learning about what others were going through helped them relate to and understand others. One participant (P53) put it, "\textit{Felt good to hear what somebody else you don’t know is going through.}" 
% C4P2 = "Felt good to hear what somebody else you don’t know is going through."

However, participants who were asked to make a comic in response to another person's comic found the activity difficult and felt it should be optional to give. Four of these participants (40\%) felt that it was hard to respond to others' comics with only this prompt and the comic materials, and preferred to either reflect on their own story or just read other people's stories and not have to respond. Two said that it was hard to respond because they hadn't dealt with similar problems and didn't feel like they could provide responses to things they hadn't been through. One participant also said that they did not feel like they understood the other person's situation, so did not know how to help them. Two participants suggested giving people a choice in whether to respond to other people's comics, because they may not know enough about the situation to be able to provide support or they may not be in a head-space to support others - for example, when they are feeling "stressed" or "angry" (P25), or "down" (P14). 
% P25 = "stressed", "angry"
% P14 = "down"

\section{Cube Web app}\label{sec:cubewebapp}
In the following section, we use feedback from the low-fidelity prototypes to inform design of Cube, a web app that supports trauma-impacted youth with emotion processing and social connection through expressive and narrative therapy techniques. We discuss the app design, implementation, and evaluation via qualitative analysis of participant feedback as well as an empathetic relations analysis between comics created by users.

\begin{figure}[t]
    \centering
    \begin{minipage}[b]{0.32\textwidth}
        \vspace{0pt}
        \includegraphics[width=\textwidth]{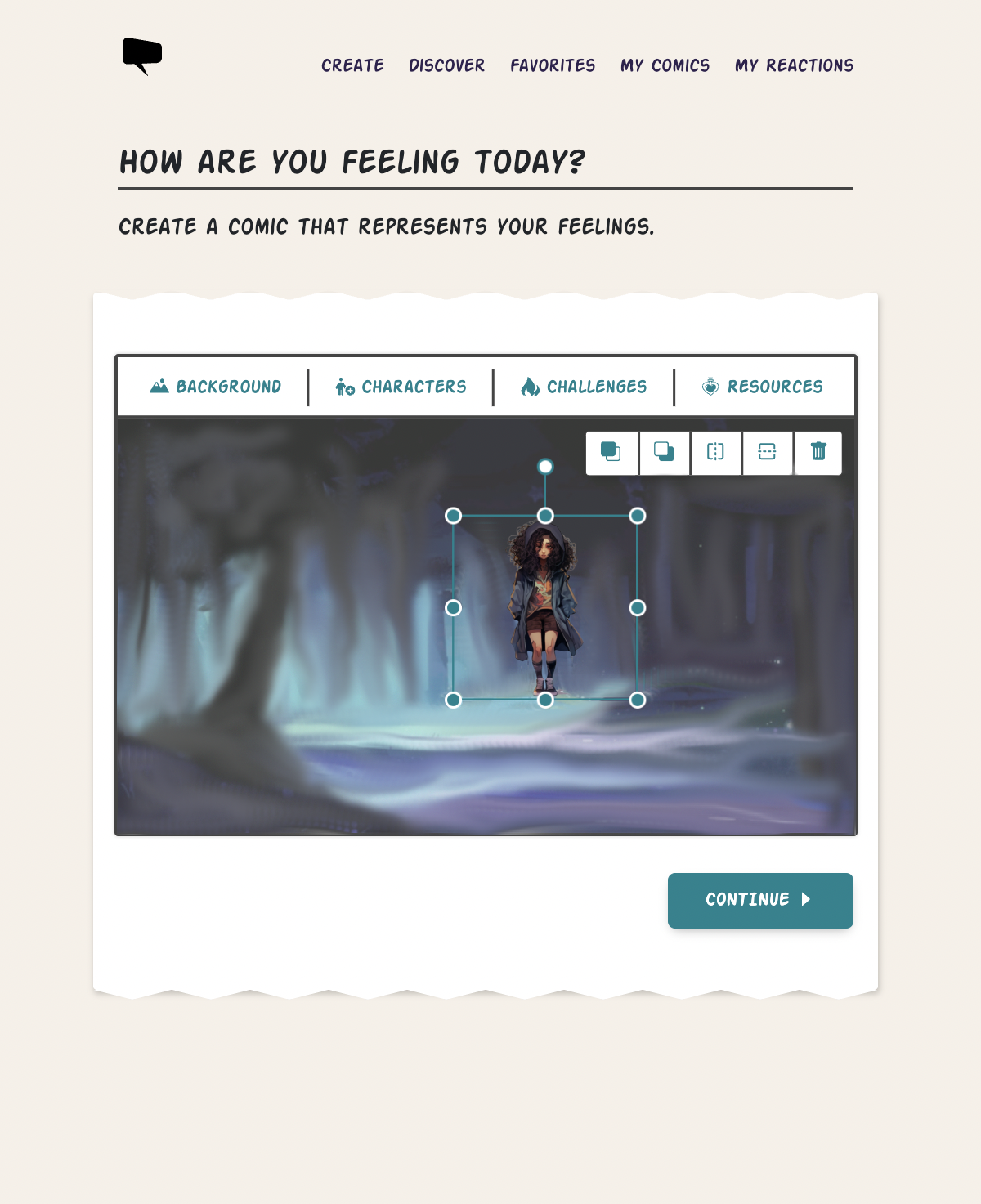}
        \caption{Users create comics in response to a chosen prompt, using supplied visual assets.}
        \Description{A depiction of a web app screen showing an interface for selecting assets from a menu and editing them within a comic panel.}
        \label{fig:createcomic}
    \end{minipage}
    \hfill
    \begin{minipage}[b]{0.32\textwidth}
        \vspace{0pt}
        \includegraphics[width=\textwidth]{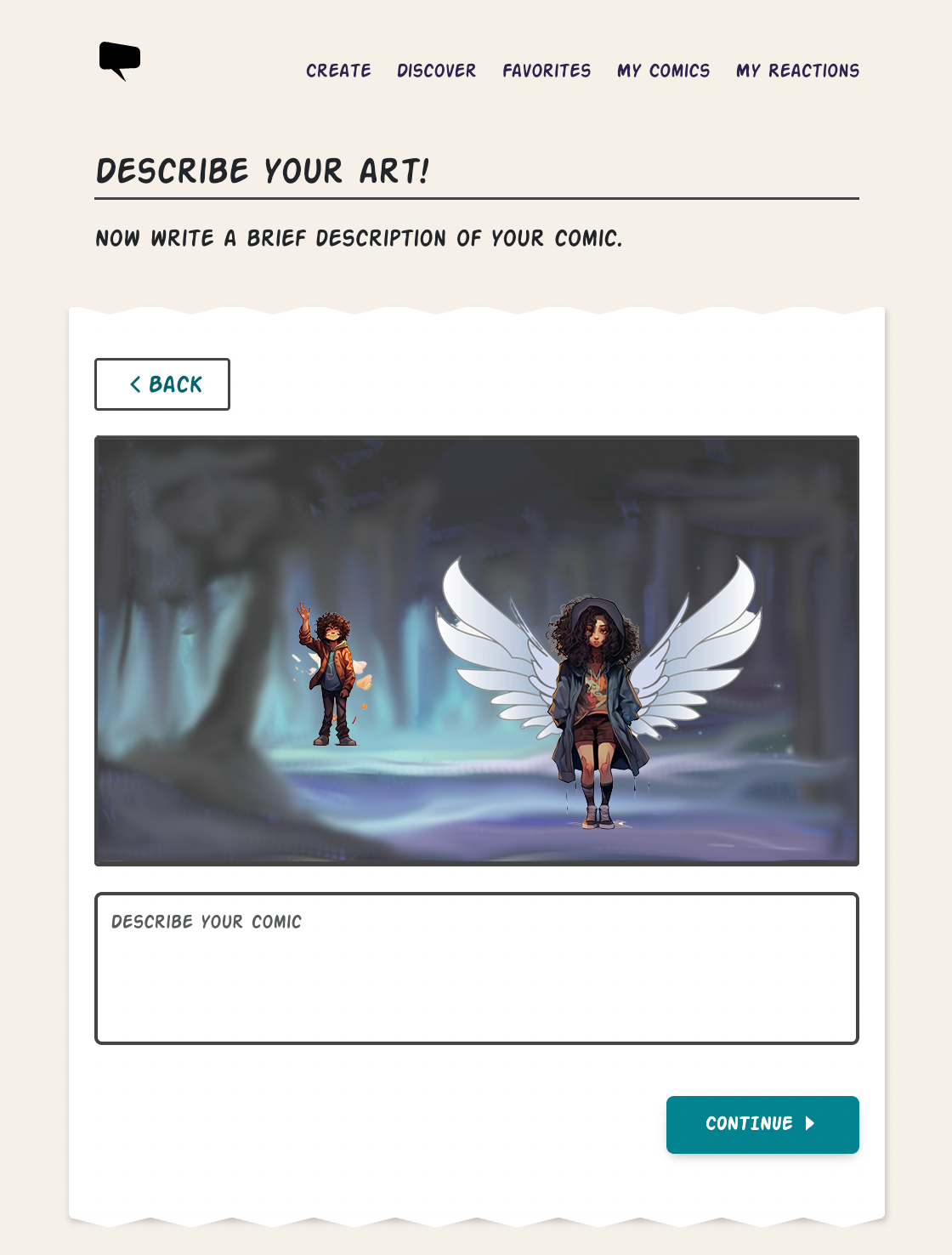}
        \caption{Users are prompted to describe their comic creations.}
        \Description{A depiction of a web app screen that shows a completed comic along with a free-response text box and message asking users to write a brief description of their comic.}
        \label{fig:describecomic}
    \end{minipage}
    \hfill
    \centering
    \begin{minipage}[b]{0.32\textwidth}
        \vspace{0pt}
        \includegraphics[width=\textwidth]{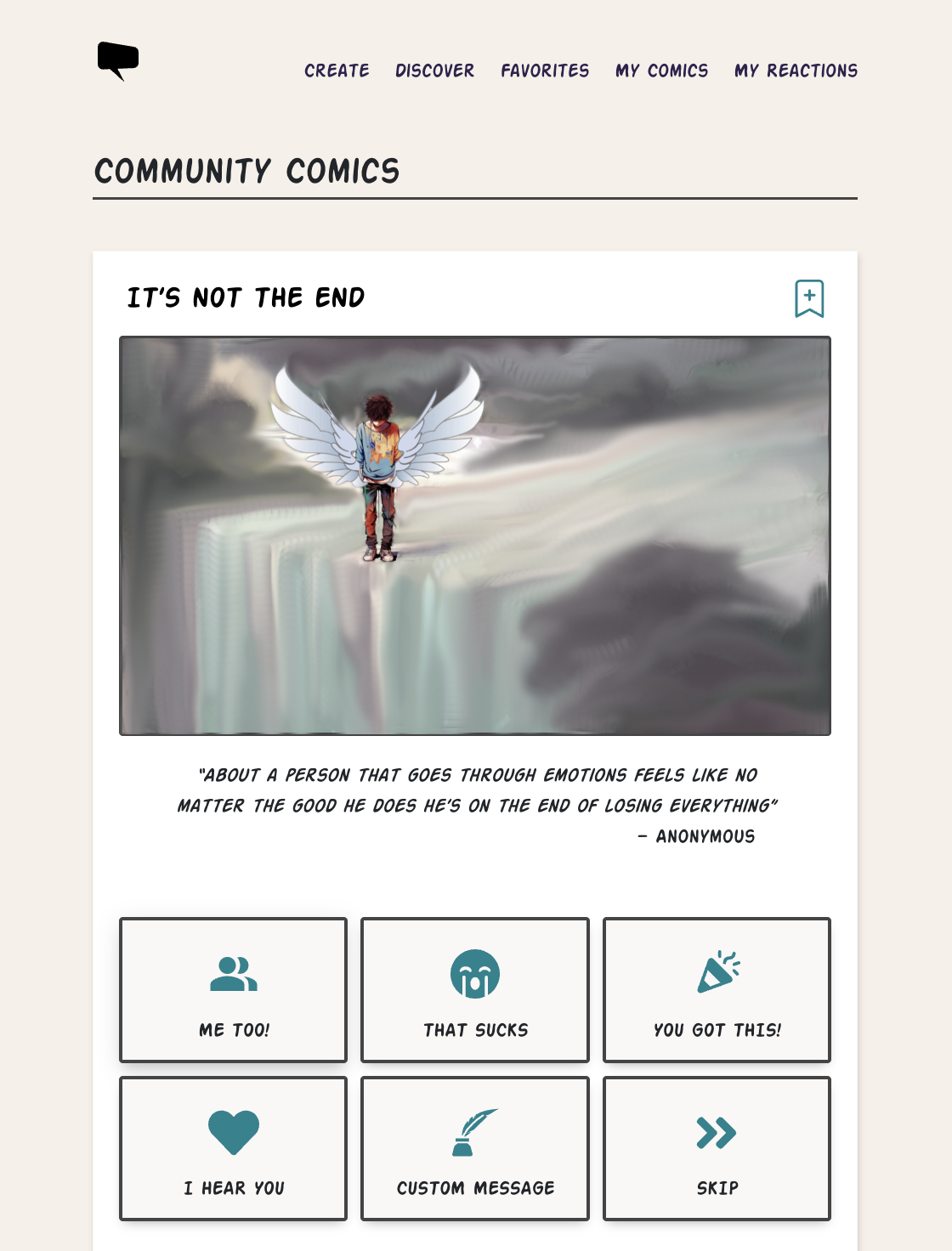}
        \caption{Users can read and respond to others' comics.}
        \Description{A depiction of a web app screen showing one comic, with options to react or skip below.}
        \label{fig:communitypage}
    \end{minipage}
\end{figure}

\subsection{App Design}

We developed Cube, a web app based on feedback from our low-fidelity prototypes, which we proceeded to test in six subsequent workshops. Based on our aforementioned results, the majority of participants found the prototypes helpful for understanding and expressing emotions, feeling more calm when experiencing strong negative emotions, and feeling less alone and more connected to others, we kept the core functionality of the app the same -- using visual assets to visually represent emotional experiences and having the ability to share these in a community of similar others. As requested by many participants, we provided more visual assets for users to choose from, especially more emotionally expressive characters and more backgrounds (36 characters, 14 backgrounds, 28 props). Given that some participants found the app too time and energy intensive and some pointed to the culprit being the requirement to come up with four-panel storylines, we shifted the app to ask users to express their feelings and experiences through single panel comics. We hypothesized that this would be less difficult and time-intensive for participants to create, while still leveraging expressive and narrative therapy techniques to facilitate the expression and processing of complex emotions and events. 

To better scaffold the comic creation process, users were prompted to create the comic visual first, and then were prompted to add a description. This was designed based on art and narrative therapeutic processes that include visual expression and verbalization, to help users and others better understand the experiences and emotions represented in their comics \cite{malchiodi_expressive_2005, white_narrative_1990}. Finally, based on the prototype-testing feedback that creating comics in response to others' comics was challenging and should not be required, we removed this activity and instead provided optional reactions that users could use to respond to others, in addition to the ability to leave a personalized message. We hypothesized that this would help participants more easily express resonance with one another, increasing overall feelings of social connectedness on the app. 

The first screen of Cube asked users to select a self-reflective comic prompt, to ensure that they were focusing on a topic that felt comfortable to them. Then, the app redirected users to an interface that allowed them to create a single-panel comic in response to the prompt, using supplied visual assets (Figure \ref{fig:createcomic}). Users could see all assets visually separated into a four category menu bar: Background, Characters, Challenges, and Resources. These assets were shown without written labels to encourage users to project onto them without verbal translation of their thoughts or feelings, as this can allow for the expression of more difficult or complex emotions \cite{malchiodi_creative_2014}. Supplied assets included background stills from an animated children's show; backgrounds, characters and objects created by digital artists; and characters and objects generated by the research team using DALLE-2\footnote{\url{https://openai.com/dall-e-2}, last accessed October 2023}, DALLE-3\footnote{\url{https://openai.com/dall-e-3}, last accessed October 2023}, Midjourney\footnote{\url{https://www.midjourney.com/}, last accessed October 2023}, and the Adobe Creative Suite\footnote{\url{https://www.adobe.com/}, last accessed October 2023}. Assets were chosen based on their ability to represent diverse experiences and people, potential to symbolize a variety of emotions, and adherence to a cohesive, magical style. Users could select a Background asset and an unlimited number of assets in other categories by selecting the item in the menu. Once selected, the asset would populate in the comic scene and could be resized, flipped horizontally or vertically, moved ahead or behind other assets in the scene, or deleted. Users could alter an asset at any time during the comic creation process by selecting it in the scene. Once users completed their comics, the app prompted them to describe their creation (Figure \ref{fig:describecomic}). In the first workshop, participants were asked to record their description with audio, as we thought the tone of voice might facilitate emotional expression. Participants, however, said this feature was highly undesirable and disruptive to the experience of using the tool. We thus modified the app for subsequent workshops to ask for a written description instead. 

Users could interact with a Discover page to read and react to other attendees' comics (Figure \ref{fig:communitypage}). This page showed users new comics one at a time and asked them to react to see the next comic, to promote user interaction. The reaction options on the app (including "Me too", "That sucks", "You got this!", "I hear you", and a custom message) were selected to make it easier for youth to express care and feelings of relatedness. We also implemented a "Skip" option for users who did not resonate with or want to respond to the comic being shown. Users could "Favorite" any comics they particularly liked, which would then appear on the Favorites page. Finally, we implemented a ``My Comics'' page where users could see all comics they created and a ``My Reactions'' page where they could see any reactions to their comics from other users.

\subsection{App Feedback}

\paragraph{Emotion Processing and Expression}

Fifteen of the thirty-four participants who interacted with the Cube app (44\%, 9 in workshop discussions and 6 via follow-up survey) explicitly stated that creating comics via the tool helped them express their emotions, confirming that Cube retained the benefits of the low-fidelity prototypes. Participants explained that the comic medium helped them express themselves in new ways, with one (P35) saying, "\textit{there is something about art imitating life}" which makes it "\textit{easy for some to express themselves through art.}" Another participant (P31) elaborated that creating a comic helped them better understand their feelings, explaining that the process helped them realize they were sad about their brother's incarceration, "\textit{I didn't even know that it was something that was bothering me}". 
% P35 = "there is something about art imitating life" which makes it "easy for some to express themselves through art."
% P31 = "I didn't even know that it was something that was bothering me"

The Cube app feedback also suggests that this benefit was not just because the tool allowed for visual creation, but because the supplied assets facilitated the process of emotional reflection and expression. One participant (P29) highlighted, "\textit{it [was] very important to have those representations of darker emotions like I saw the hole... the wall, and the fire.}" Furthermore, in 36\% of comic descriptions ($N=15$, out of the 42 comics created on Cube), participants used the assets to help verbalize their emotions. For example, one participant wrote, "\textit{...crying, anxious, feeling trapped in a bubble of my emotions while another wave of emotions is coming}", which referenced the assets in their comic, which included a boy encased in a bubble next to a big approaching wave. Participants also talked about feeling "surrounded by walls" and followed by a "dark cloud", mirroring their use of brick wall and dark cloud assets (respectively). This suggests that the assets helped young people conceptualize and/or articulate their feelings. 
% P29 = "it [was] very important to have those representations of darker emotions like I saw the hole... the wall, and the fire."
% [P_07cc4305-4712-8aa7-95da-c2f1451dd560] = "...crying, anxious, feeling trapped in a bubble of my emotions while another wave of emotions is coming"
% [P_a07a62cb-38f1-912c-5996-5d109159fb4b] = "surrounded by walls"
% [P_64b5b2df-7cd8-1f4a-f2cd-4191981906bf] = "dark cloud"

Despite the team's efforts to create a more comprehensive set of assets, twelve participants (35\%, 7 in workshop discussions and 5 via follow-up survey) still advocated for more supplied assets, with the majority of these participants ($N=10$) requesting more assets of all types (backgrounds, characters, obstacles, and resources) so they could depict more subjects or find assets they resonated with more fully. Three participants specifically asked for "\textit{more characters that look like us}" (P50), suggesting the app include a more diverse character set or allow users to customize their characters. Three participants also talked about wanting more depictions of emotion, including more expressive characters and backgrounds and objects that can be used to create emotional metaphors. Like in the prototype-testing feedback, no participants asked for the tool to include an infinite number of assets, but rather a more comprehensive, but limited, set for users to choose from. 
% FoC\_Youth = "more characters that look like us"

Six participants (18\%, 5 in workshop discussions and 1 via follow-up survey) indicated that they did not like the comic format in general, advocating for being able to express themselves verbally rather than visually. For example, one participant (P31) stated, "\textit{I’m in between liking it and not liking it... since I’m more verbal... The visual representation was nice, but I would not really want to make anything since I'm not really a crafty person.}" Another participant (P51) explained that they don't want to express themselves through art because they "\textit{can't make it look unique and good [in] under 5 minutes}" and in moments of high emotional intensity, would rather write or use voice chat to express themselves. However, no participants brought up concerns about creating comics with Cube requiring too much time or energy, which had been an issue for those who interacted with the low-fidelity prototypes. 
% P31 = "I’m in between liking it and not liking it... since I’m more verbal... The visual representation was nice, but I would not really want to make anything since I'm not really a crafty person."
% P51 = "can't make it look unique and good [in] under 5 minutes"

% Furthermore, participants appeared to enjoy the process of creating comics, with six describing how it was fun to creatively express themselves and create something visually appealing. One participant (SFLA\_P5) explained that this had a positive impact on their mood, stating that "making a comic scene about my emotions was fun and made me forget about my problems for a bit."

\paragraph{Sense of Connection}

Cube, like the low-fidelity prototypes, appeared to facilitate social connection. Nine participants (26\%, 4 in workshop discussions and 5 via follow-up survey) shared that they could relate to the emotions or challenges described in other people's comics. For example, in response to a comic, one participant (P26) explained, "\textit{This describes my life right now,}" citing that like the author, they were currently "\textit{working with what [they] have}" and "\textit{learning how to deal with [challenges] as an adult.}" Another participant (P38) elaborated that the medium helped them connect with others' comics, saying that they related to the comics they read because "\textit{a lot of the people were expressing themselves quite vividly}" and they could "\textit{feel the sadness and frustration through both the comics and the words}." Six participants (18\%, 2 in workshop discussions and 4 via follow-up survey) added that the emotional resonance they felt with others' comics helped them feel less alone and more supported. The previously quoted participant (P38) explained, "\textit{[Reading others' comics] helped me feel more supported, I feel at times I like to complain too much and my feelings are not valid. I try to keep them to myself so I don’t seem like a negative or miserable person but it feels good to know I am not alone and that many people also feel the same}".
% P26 = "This describes my life right now", citing that like the author, they were currently "working with what [they] have" and "learning how to deal with [challenges] as an adult."
% P38 = "a lot of the people were expressing themselves quite vividly" and they could "feel the sadness and frustration through both the comics and the words."
% P38 = "[Reading others' comics] helped me feel more supported, I feel at times I like to complain too much and my feelings are not valid. I try to keep them to myself so I don’t seem like a negative or miserable person but it feels good to know I am not alone and that many people also feel the same".

In contrast to how challenging participants found it to respond to one another in the low-fidelity prototype system, five participants (15\%, 2 in workshop discussions and 3 via follow-up survey) talked about how Cube's response options made it feel quick and easy to express feelings of connection or care for other users. One participant (P29) explained that having easy options for responding allows young people to "\textit{pick and choose [how] we want to show up and relate to people}" in a way that "\textit{doesn't feel overwhelming.}" Another participant (P38) added, "\textit{I think the response options are quite direct and it makes it easy to respond to the comics. I like the different options it has rather then just a like or dislike button [because] it allow space for interactions.}"   
% P29 = having multiple easy options for interaction allows young people to "pick and choose [how] we want to show up and relate to people" in a way that "doesn't feel overwhelming."
% P38 = "I think the response options are quite direct and it makes it easy to respond to the comics. I like the different options it has rather then just a like or dislike button [because] it allow space for interactions."

At the same time, four participants (12\%, 2 in workshop discussions and 2 via follow-up survey) had suggestions on how the reaction options could be improved. In particular, these participants felt that the "That sucks" reaction should be changed, with one (P33) explaining, "\textit{I initially interpreted 'that sucks' as an insult.}" They also advocated for allowing users to customize the reactions before sending them out, since "\textit{using [our] slang makes it more personal}" and helps build "a sense of community" (P29).

Four participants (12\%, 2 in workshop discussions and 2 via follow-up survey) wanted to be able to connect more deeply with other users, either via public comments or private messages. One participant (P29) elaborated that these options would allow users to "\textit{say something more personal}" and express sentiments like, "\textit{I really relate to you. I really can feel what you're going through. I wanna talk to you, make sure you're okay.}" 

Three participants (9\%, 2 in workshop discussions and 1 via follow-up survey) also talked about the importance of protecting users' safety and privacy in these community interactions, appreciating that the app's current anonymity "\textit{makes people not feel embarrassed of having a hard time}" (P38), and suggesting the addition of community guidelines, an option to turn off private messages, and a monitoring system to ensure that custom messages exchanged on the platform are appropriate. 
% P33 = "I initially interpreted 'that sucks' as an insult."
% P29 = "using [our] slang makes it more personal" and helps build "a sense of community" 
% P29 = "say something more personal" and express sentiments like, "I really relate to you. I really can feel what you're going through. I wanna talk to you, make sure you're okay."
% P38 = "makes people not feel embarrassed of having a hard time"

%%5 participants generally liked being able to see other people's comics, citing that they enjoyed seeing the creative ways others expressed themselves and learning about what others were experiencing. 
%%- How does it feel to look at other people’s comics?  ToU7pm_P5: It's an eye/mind opening experience just knowing people feel the same or have different thoughts and it shows how they go about expressing themselves. [P2 and P1 agrees]
%%- How do you feel about the sharing part?  ToU1pm_P1: it was cool to see their creativity and what other people are going through; I thought that was really nice

\begin{figure}[t!]
    \centering
    \includegraphics[width=0.8\linewidth]{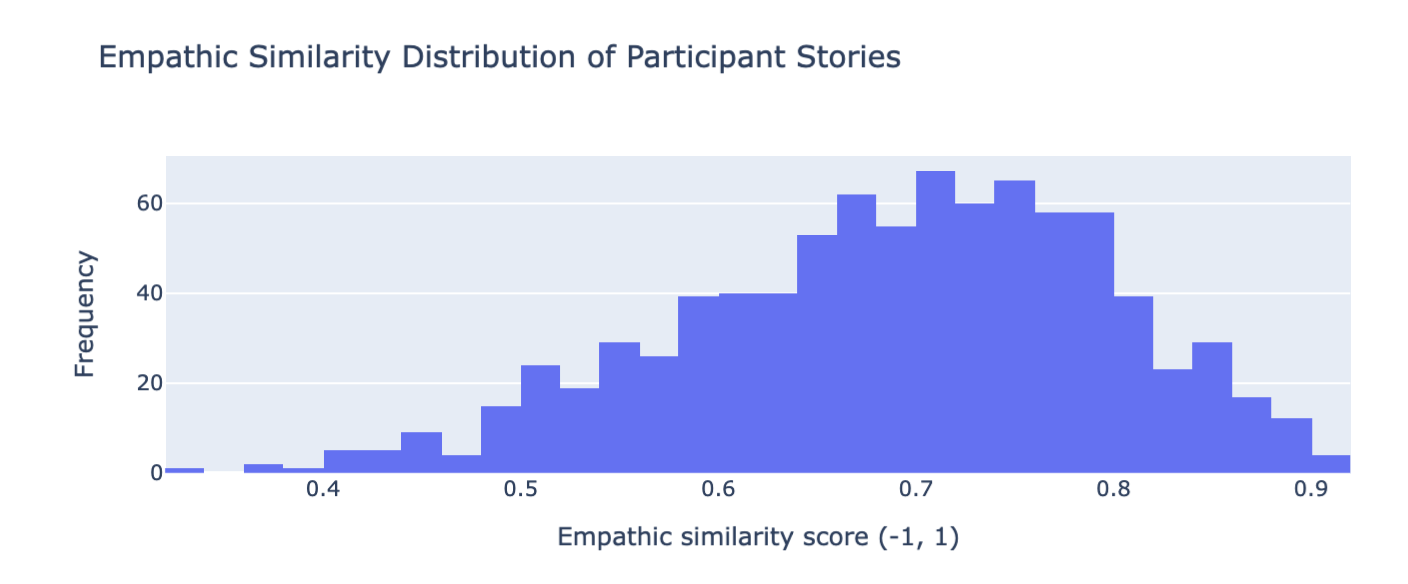}
    \caption{Distribution of empathic similarity between all pairs of comics shared. Empathic similarity captures the level two narrators are likely to empathize with one another given two pieces of text.}
    \Description{A histogram of empathic similarity scores, which range from -1 to 1. The histogram is skewed left, and the peak is between scores of 0.7-0.8.}
    \label{fig:distr}
\end{figure}
\begin{figure*}[t!]
    \centering
    \includegraphics[width=.8\textwidth]{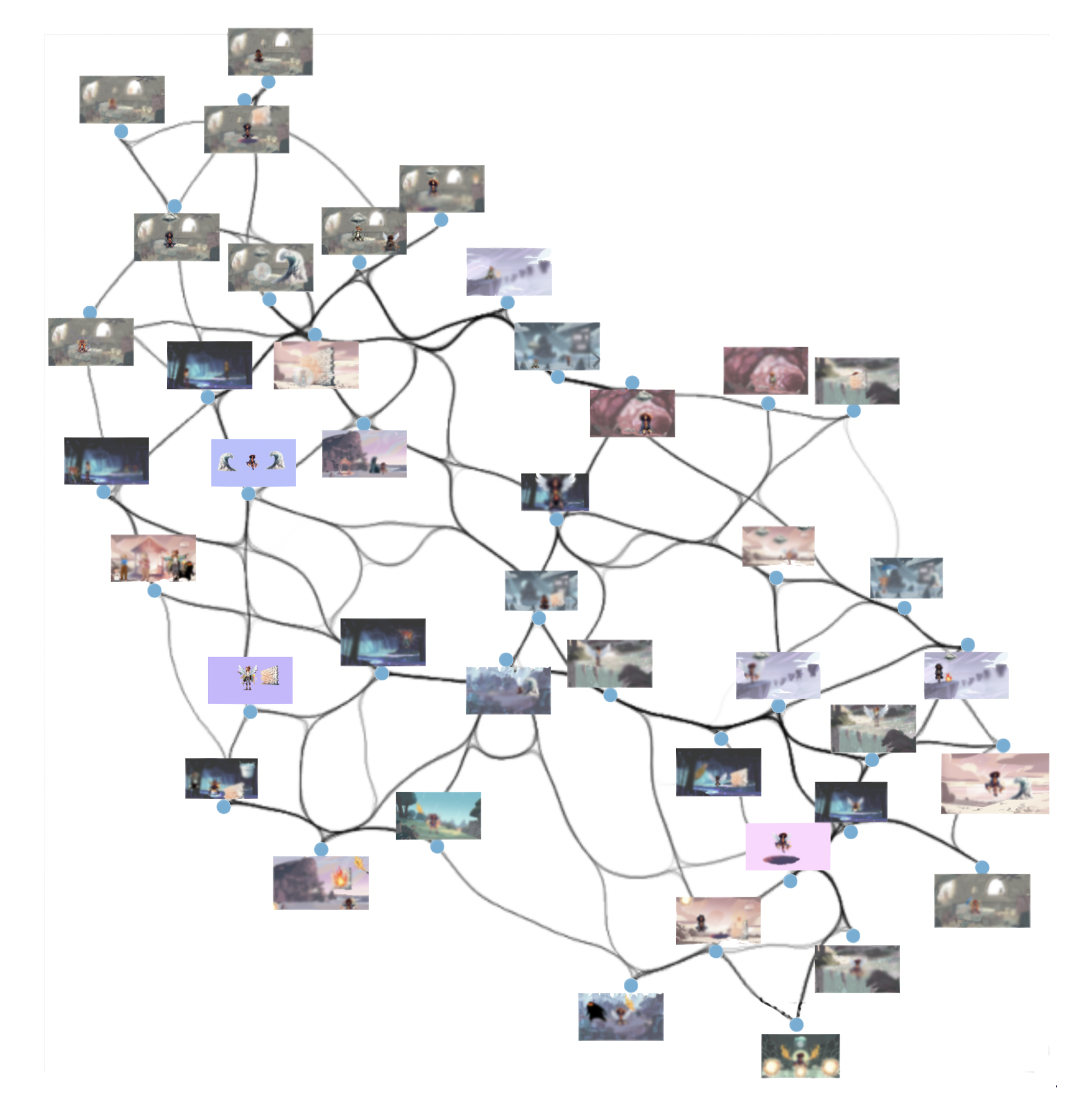}
    \caption{Visualization of all comic embeddings in 2D space (using UMAP) with comics overlayed. Black lines indicate the weight of connectivity between comics, where darker lines indicate higher weight.}
    \Description{A visualization of all participant comics laid out in a web, spaced out depending on their level of empathic similarity to one another. In general, comics with corresponding background images and characters are grouped closer together.}
    \label{fig:comicweb}
\end{figure*}

\subsection{Empathetic Relationship Between Comics}
In the following section, we use quantitative analysis to explore whether participants can empathize with each others' comics and understand what features contribute to empathetically similar comic clusters. Recall from Section  \ref{sec:dataanalysis} that we use the model from \citet{shen_modeling_2023}, which was trained on another narrative data set to evaluate ``empathetic similarity'' between two stories. These scores indicate the degree to which the narrators of 2 stories are likely to empathize with one another, and range from -1 to 1, where -1 means the two narrators would not empathize with one another at all and 1 means the two narrators would empathize with one another completely. 

Figure \ref{fig:distr} shows the distribution of empathetic similarity between all the pairs of comics shared by participants. As we can see from the distribution, the comics that participants shared are generally empathetically similar to one another (the distribution is left-skewed), with a mean empathic similarity of $0.69$ and standard deviation of $0.11$. On average, participants' comics are positively empathetically related, indicating that if participants read one anothers' comics, they would generally be able to resonate with them.

\begin{figure*}[t!]
    \centering
    \includegraphics[width=.8\textwidth]{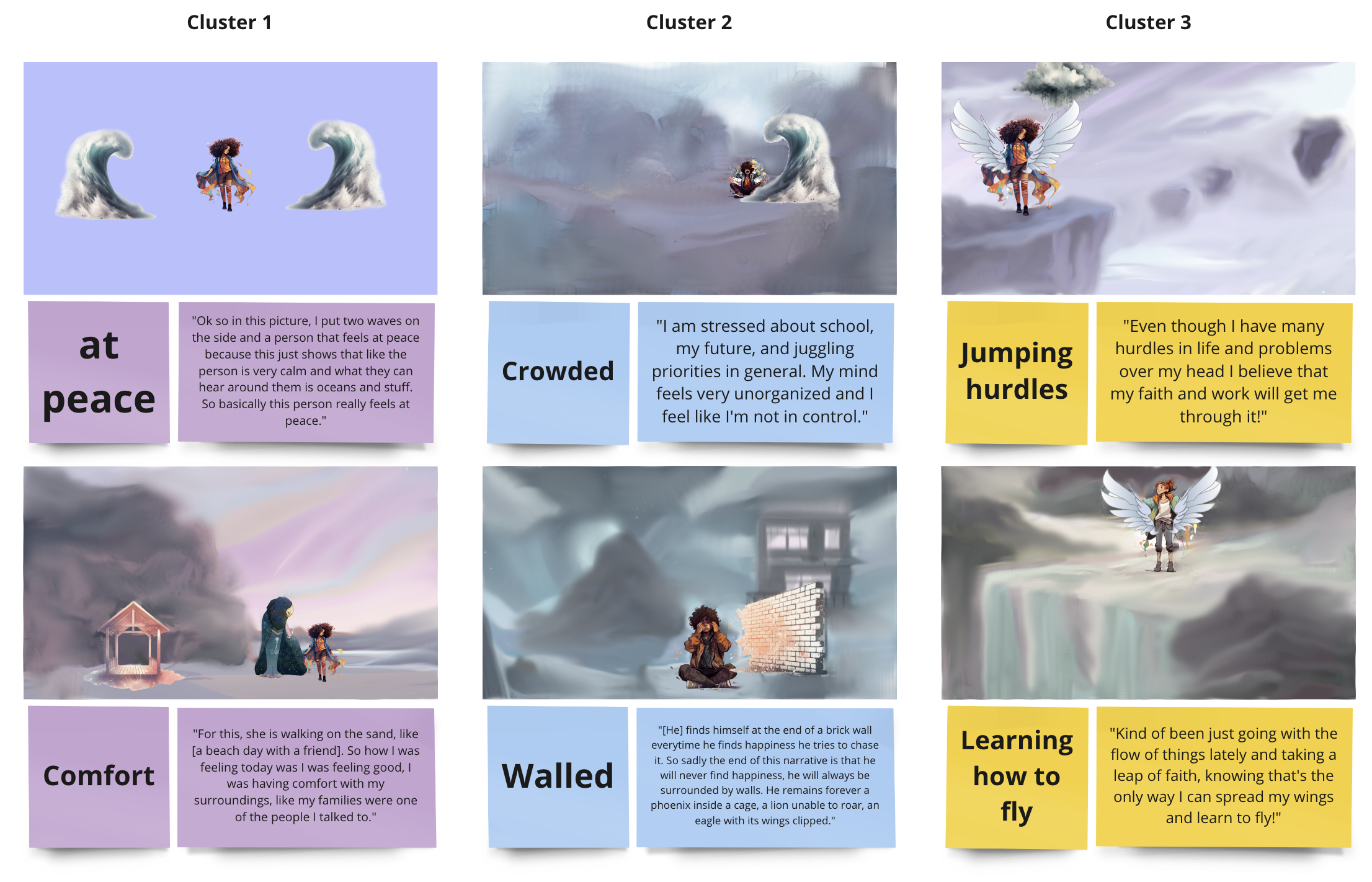}
    \caption{Examples of clusters with high empathetic relationship between the comics. Titles and captions of the comics are shown in sticky notes below each comic.}
    \Description{An image with 3 clusters showing pairs of stories within each cluster that have high empathetic similarity. Cluster 1 shows two comics describing the feelings of peace and comfort -- Story 1: "Ok so in this picture, I put two waves on the side and a person that feels at peace because this just shows that like the person is very calm and what they can hear around them is oceans and stuff. So basically this person really feels at peace.", Story 2: "For this, she is walking on the sand, like [a beach day with a friend]. So how I was feeling today was I was feeling good, I was having comfort with my surroundings, like my families were one of the people I talked to." Cluster 2 shows two comics describing the feelings of stress and anxiety -- Story 1: "[He] finds himself at the end of a brick wall everytime he finds happiness he tries to chase it. So sadly the end of this narrative is that he will never find happiness, he will always be surrounded by walls. He remains forever a phoenix inside a cage, a lion unable to roar, an eagle with its wings clipped", Story 2: "I am stressed about school, my future, and juggling priorities in general. My mind feels very unorganized and I feel like I'm not in control." Finally, cluster 3 shows two comics describing the feelings of overcoming -- Story 1: "Even though I have many hurdles in life and problems over my head I believe that my faith and work will get me through it!", Story 2: "Kind of been just going with the flow of things lately and taking a leap of faith, knowing that‚ the only way I can spread my wings and learn to fly!" These examples show that similar emotion pairs are visually very different.}
    \label{fig:clusters}
\end{figure*}

Next, we explore how visual features of the comics created by participants influenced connections towards other comics. This allows us to better understand how the Cube app features allowed users to empathize more or less with other peoples' comics. 
Figure \ref{fig:comicweb} shows a visualization of the comics embedding in 2D space, with the participants' comics overlayed and the strength of the connectivity between comics shown in black. In general, comics with similar overall color and mood are clustered together, or marked as empathetically similar, however, there are exceptions towards the middle of the visualization. In particular, clusters exhibit similarities in (1) environments (gray room, mystical forest, above pink clouds), (2) character gestures (ie. arms spread far apart, sitting, or dancing), and (3) symbols (ie. wings, walls, waves). Figure \ref{fig:clusters} shows examples of pairs with high empathic similarity. These reveal that emotions are conveyed in multi-modal ways using colors, gestures, and symbolism of the image as well as the overall emotion of the stories. 

We can see from Figure \ref{fig:comicweb} that empathetically similar comics are not grouped by similar visuals alone, and that the comic descriptions play a role in the relationship between comics. The examples in Figure \ref{fig:clusters} show how different images can be used to convey a shared emotion, such as feeling tranquil, feeling trapped, or overcoming barriers. For example in Cluster 3, one participant's comic, "jumping hurdles," shows the character amidst stepping stones in the sky, while another participant's comic, "learning how to fly" depicts a character at the edge of a waterfall. Despite differences in the characters and backdrops, the feelings conveyed in the comics appears similar, as suggested by their descriptions (``...I believe that my faith and work will get me through it!" and "...I can spread my wings and learn to fly").

\section{Discussion and Design Implications}
% Metaphors and symbols are present not only in the images, but also in the text (ie. in cluster 2, a participant shared, \textit{``He remains forever a phoenix inside a cage, a lion unable to roar, an eagle with its wings clipped''})

\paragraph{Emotion Processing and Expression}
In this work, we found that Cube can help youth and young adults with emotion processing and self-expression. In workshops evaluating low fidelity prototypes, we saw that the comic medium helps young people understand how they are feeling and can help them calm down when experiencing anger, sadness, or anxiety. These results are consistent with prior literature on expressive therapy, and could be explained as a grounding or anchoring effect, which helps patients draw attention to the present moment \cite{tucci_handbook_2019, haeyen_validity_2021}. As indicated by feedback and artifacts from workshops evaluating the Cube app, the medium appeared to help youth express themselves in new and creative ways, such as conceptualizing or verbalizing their feelings in both visual and symbolic ways. Additionally, the scaffolding of the visual tool seemed to help with the projection and externalization emotions, much like methods such as sandtray therapy \cite{malchiodi_expressive_2005, sweeney_sandtray_2021}. From quantitatively analyzing empathetic content across the comics, we saw that the ways participants visually expressed their emotions were different, even for similar emotions. These results emphasize the importance of providing participants with many visualization options to facilitate their full and nuanced self-expression. 

We found from evaluations of both the low-fidelity prototypes and the Cube web app that participants desired more assets for comic creation. In particular, participants felt that more assets and diversity in assets, without endless options, would allow them to identify with characters better and create more resonant stories without getting overwhelmed. They expressed a desire for more depictions of emotion, such as through expressive characters and backgrounds, as well as more objects to create emotional metaphors. Future work can explore the balance of providing participants with many creative tools without the cognitive load of too many options. %% during an already difficult emotion processing task.

A subset of participants stated that they preferred expressing themselves verbally rather than visually. Some prior works suggest that a combination of both verbal and visual expression can enhance the efficacy of expressive therapies, and that verbal and nonverbal behaviors in relation to emotion are dependent on the type of emotion discussed \cite{lee_effect_2002, berry_nonverbal_2010}. As such, it is important for future designs to consider personalizing multimedia expression options based on individual preferences and the user's context. Furthermore, multiple participants who said they were more verbal than visual elaborated that they were not crafty or that they could not create a good comic. In the future, we recommend that researchers compare the impact of a written-only activity to that of a visual one, investigating if particular subsets of the population are better suited to one medium or another. Future work could also explore how social comparison regarding sharing of the comics may affect participants' confidence as creators of their own visual stories.

\paragraph{Social Connection}
We found that reading other participants' comics can help young people feel less alone and more connected to the experiences of others. In workshop feedback, participants shared that they could relate to the emotions and challenges described in other people's comics, amplified by the visuals \cite{silvia_emotional_2005}. The quantitative empathetic content analysis indicated that comics had a high degree of empathic similarity, suggesting that participants would be able to relate to each others' comics if the system were launched at scale. In addition, we qualitatively show through visualizing the comics that the artistic elements provided as features through Cube contribute to empathy towards the stories: for example, participants use color, environment, and symbol assets to convey resonant messages such as ``taking a leap of faith'' or ``jumping over hurdles.'' These results suggest the power of the expressive art features in relaying personal and relatable messages to other youth who have experienced developmental trauma. 
Future designs might explore if there are benefits to amplifying this effect by giving users opportunities to collaboratively reflect on user's narratives, such as through smaller groups in the app, in order to enhance users' sense of connection while preserving safety.

From prototype feedback, we learn that instructing participants to construct comics in response to others was difficult. Participants shared that they did not feel like they knew enough about the user's story to support them, or felt that they were not in the right headspace to respond. This phenomenon is in line with research on social overload, which can occur on social platforms when users feel they are giving too much support to others \cite{maier_giving_2015}. In iterating on the design for Cube, we learned that participants appreciated the scaffolding response options we constructed, which made it much quicker and easier to express their feelings of connection or care for others. They additionally requested greater flexibility in the wording of the responses. Future designs could continue to scaffold the process of providing social support, while allowing some customization of supportive messaging.

Finally, although no participants reported feeling unsafe using any version of the tool and only a small percentage mentioned safety considerations as being important to its future development, it is important for future work to further investigate the safety risks of this type of platform and identify how best to mitigate them prior to launching with a broader audience. For instance, future work may explore different aspects of the emotional impact of our tool, such as if exposure to many negative emotions can cause emotion fatigue in users.
% if we're going to deploy this without supervision online?) or mention that is future work and the focus here is within a small group of people of similar age range?

\paragraph{Design Insights}

Based on the findings of this study, we propose the following high-level design recommendations for future interventions that aim to increase emotional expression and social connection for young people who have experienced trauma.
\begin{enumerate}
    \item \textbf{Visual expression of emotional experiences.} 
In our study, we saw that scaffolded visual storytelling frequently allowed young people to easily express emotions in new and creative ways, which is in line with works demonstrating the efficacy of expressive and narrative-based therapies \cite{malchiodi_creative_2014, sweeney_sandtray_2021, white_narrative_1990}. While youth advocated for additional assets to be added to the system, they also appreciated that the platform made it easy to create images that represented complex feelings. It is important for future designs to focus on balancing having enough visualization options to allow for creative expression while also limiting the scope enough for the task to not be overwhelming or overly time-consuming, especially for individuals who do not identify as artists. 
    \item \textbf{Sharing in a community of peers.}
Our results highlight the value of building systems through which youth can share their visual expressions with others who have similar experiences. In our study, we find that seeing the visuals of others has the potential to help young people feel less alone and more connected because they can relate to the emotional experiences portrayed in others' comics. This effect is present even when youth do not interact with one another directly, which is tied to prior literature emphasizing the need for enhancing ``diffuse sociality,'' or the general feeling of connection towards others regardless of direct interaction \cite{bhattacharjee_i_2022-1}. This type of sociality, related to constructs of compassionate love towards humanity in social psychology \cite{sprecher_compassionate_2005}, may be particularly beneficial for connecting youth who have experienced developmental trauma, as direct sociality could be difficult or uncomfortable. 
    \item \textbf{Support for peer-to-peer responses.}
In our study, youth both expressed that templated response options made it easier for them to express connection or care for others and had a desire to customize these options prior to usage. These findings build off of prior literature suggesting templates can help users navigate digital communities \cite{zhang_separate_2022}. We recommend that future designs look for ways to support youth in expressing feelings of resonance easily yet in their own voice, to encourage supportive online interactions.  
\end{enumerate}
    
\paragraph{Broader Implications for Mental Well-being}
In this work, we conducted an iterative design process with youth who have experienced developmental trauma. Our app design is inspired by interdisciplinary approaches related to expressive and narrative therapies as well as social technology design. As such, it is important to understand how trauma-impacted youth may benefit from our technology in therapeutic ways. 

In particular, we envision our work being integrated with existing online platforms and services used by trauma-impacted youth, improving mental well-being of such youth via the following therapeutic veins: Firstly, in the comic creation process, youth are able to visually externalize their internal experiences, aiding in emotion expression via projection \cite{malchiodi_expressive_2005}. This comic format may be easier for people who have experienced deep trauma, and who are just beginning to be able to share any of their experience or related feelings (in contrast to other cases, where a person may be comfortable engaging in a deep written retelling of their personal story). After visually representing their emotions, putting feelings into words via the comic description is an important avenue for experience processing \cite{malchiodi_expressive_2005} and implicit emotion regulation \cite{torre_putting_2018}, and can distance users from the difficult, trauma-filled stories they share \cite{white_narrative_1990}.
The creation process may also allow youth to enter a state of creative flow or mindful focus. Creativity and mental well-being are tied to the flow experience, which can be achieved during art making \cite{chilton_art_2013}, and is well-known to improve both short term mood and long-term well-being. For example, one participant expressed ``\textit{making a comic scene about my emotions was fun and made me forget about my problems for a bit.}'' 

Secondly, through seeing the comics shared by others who have had similar experiences - either in a stand-alone system like Cube, by sharing comics on existing digital platforms that trauma-impacted youth visit (such as resource-focused apps for foster-involved youth), or by sharing comics within human-facilitated programs for trauma-impacted youth (such as group therapy or skill-building programs) - youth may feel more connected and less alone in their situations. Receiving positive feedback from peers can be facilitated by templated responses, which we envision being integrated into existing social platforms for trauma-impacted youth. This scaffolded response process enabled by technology can provide further social support and increase youths' confidence in expressing their emotions in general, opening them up to forming healing and protective social connections in their everyday lives. This increased social support, here across digital and physical domains, is critical for youths' ability to heal from trauma and achieve mental well-being \cite{cohen_stress_1985,herman_trauma_2015,kawachi_social_2001}. 

% integrating with existing social platforms

% therapeutic veins:
% 1. in the comic creation process, flow state, and measuring emotion processing or emotion awareness
% [flow state + externalization/projection] % where we can add quotes / related work

% 2. in the comic sharing process (group therapy, social support, empathy and interaction)

\paragraph{Limitations and Future Work}
While our work offers new insights regarding the design of social platforms for young people who have experienced trauma, there are several limitations. Because we did not ask participants to disclose their trauma histories, we do not know whether the findings are limited to a particular subset of young people who have experienced their unique traumas. While we hosted numerous smaller workshops over time and recruited youth and young adults from organizations in multiple geographic regions to enhance diversity and reduce group think, it's possible that the workshop feedback could be subject to group think, rather than obtaining fully individualized responses. Additionally, since participants were using the tool within the context of a workshop rather than asynchronously (as is intended for real-world usage), further research is needed to determine how youth feel about the tool when using it in their homes and everyday lives. We did attempt to mitigate this limitation by ensuring that all participants used the tool from their own personal devices. Due to the length of the workshops and limited sample size, it was not possible to have all comics read and rated by other participants, which limited our evaluation of empathetic similarity.  Similarity was thus computed using a pre-trained model, which may have led to different results than if comic resonance was formally examined by youth. We will aim in future work to deploy Cube in-the-wild with a larger population of young people, and gather in-the-moment as well as follow-up feedback from participants on their experiences with Cube. Finally, while our designs were based on therapeutic techniques and informally evaluated by subject matter experts during the development process of Cube, a formal assessment of efficacy as it relates to validated therapeutic interventions is needed to determine the therapeutic impact of the tool for youth who have experienced developmental trauma. 

\section{Conclusion}
In this paper, we work with young people who have experienced trauma to design and evaluate a novel digital platform, Cube. Cube uses therapeutic techniques to support processing, expressing and sharing emotions in safe and healthy ways. We iteratively improve and evaluate Cube through a series of nine in-person workshops with $N=54$ youth and young adults, where the first cohort interacted with low-fidelity prototypes and a second cohort interacted with the refined Cube web app. The app supports participants to construct comics describing their personal experiences and scaffolds them in responding to the emotional stories expressed by others. From qualitative and quantitative analyses, we find that Cube helps young people with (1) emotion processing and expression and (2) feeling connected to the shared experiences of their peer community. In addition to the creation of Cube, this work also contributes insights that may help designers who aim to create safe and supportive social technologies, especially involving vulnerable youth.

\begin{acks}
Thank you to our anonymous reviewers for their thoughtful feedback on this paper, which has greatly improved its quality. We also would like to express our gratitude to the participants who contributed their time and wisdom in this project, as well as the community organizations that supported the process. Finally, we sincerely thank the Shah Family Foundation and our other donors for providing the funding that has allowed us to embark on this project in the first place. 
\end{acks}

\bibliographystyle{ACM-Reference-Format}
\bibliography{sample-base}


\begin{thebibliography}{68}

%%% the Web address.

\ifx \showCODEN    \undefined \def \showCODEN     #1{\unskip}     \fi
\ifx \showISBNx    \undefined \def \showISBNx     #1{\unskip}     \fi
\ifx \showISBNxiii \undefined \def \showISBNxiii  #1{\unskip}     \fi
\ifx \showISSN     \undefined \def \showISSN      #1{\unskip}     \fi
\ifx \showLCCN     \undefined \def \showLCCN      #1{\unskip}     \fi
\ifx \shownote     \undefined \def \shownote      #1{#1}          \fi
\ifx \showarticletitle \undefined \def \showarticletitle #1{#1}   \fi
\ifx \showURL      \undefined \def \showURL       {\relax}        \fi
% invisible to TeX
\providecommand\bibfield[2]{#2}
\providecommand\bibinfo[2]{#2}
\providecommand\natexlab[1]{#1}
\providecommand\showeprint[2][]{arXiv:#2}

\bibitem[Andalibi et~al\mbox{.}(2018)]%
        {andalibi_social_2018}
\bibfield{author}{\bibinfo{person}{Nazanin Andalibi},
  \bibinfo{person}{Oliver~L. Haimson}, \bibinfo{person}{Munmun~De Choudhury},
  {and} \bibinfo{person}{Andrea Forte}.} \bibinfo{year}{2018}\natexlab{}.
\newblock \showarticletitle{Social {Support}, {Reciprocity}, and {Anonymity} in
  {Responses} to {Sexual} {Abuse} {Disclosures} on {Social} {Media}}.
\newblock \bibinfo{journal}{\emph{ACM Transactions on Computer-Human
  Interaction}} \bibinfo{volume}{25}, \bibinfo{number}{5} (\bibinfo{date}{Oct.}
  \bibinfo{year}{2018}), \bibinfo{pages}{1--35}.
\newblock
\showISSN{1073-0516, 1557-7325}
\href{https://doi.org/10.1145/3234942}{doi:\nolinkurl{10.1145/3234942}}


\bibitem[Andalibi et~al\mbox{.}(2022)]%
        {andalibi_lgbtq_2022}
\bibfield{author}{\bibinfo{person}{Nazanin Andalibi}, \bibinfo{person}{Ashley
  Lacombe-Duncan}, \bibinfo{person}{Lee Roosevelt}, \bibinfo{person}{Kylie
  Wojciechowski}, {and} \bibinfo{person}{Cameron Giniel}.}
  \bibinfo{year}{2022}\natexlab{}.
\newblock \showarticletitle{{LGBTQ} {Persons}’ {Use} of {Online} {Spaces} to
  {Navigate} {Conception}, {Pregnancy}, and {Pregnancy} {Loss}: {An}
  {Intersectional} {Approach}}.
\newblock \bibinfo{journal}{\emph{ACM Transactions on Computer-Human
  Interaction}} \bibinfo{volume}{29}, \bibinfo{number}{1} (\bibinfo{date}{Jan.}
  \bibinfo{year}{2022}), \bibinfo{pages}{2:1--2:46}.
\newblock
\showISSN{1073-0516}
\href{https://doi.org/10.1145/3474362}{doi:\nolinkurl{10.1145/3474362}}


\bibitem[Anderson and Gehart(2012)]%
        {anderson_collaborative_2012}
\bibfield{author}{\bibinfo{person}{Harlene Anderson} {and}
  \bibinfo{person}{Diane Gehart}.} \bibinfo{year}{2012}\natexlab{}.
\newblock \bibinfo{booktitle}{\emph{Collaborative {Therapy}: {Relationships}
  {And} {Conversations} {That} {Make} a {Difference}}}.
\newblock \bibinfo{publisher}{Routledge}.
\newblock
\showISBNx{978-1-135-92625-0}
\newblock
\shownote{Google-Books-ID: ct9ONlDG3dIC}.


\bibitem[Badillo-Urquiola(2020)]%
        {badillo-urquiola_social_2020}
\bibfield{author}{\bibinfo{person}{Karla Badillo-Urquiola}.}
  \bibinfo{year}{2020}\natexlab{}.
\newblock \showarticletitle{A {Social} {Ecological} {Approach} to {Empowering}
  {Foster} {Youth} to be {Safer} {Online}}. In
  \bibinfo{booktitle}{\emph{Companion {Publication} of the 2020 {Conf} on
  {Computer} {Supported} {Cooperative} {Work} and {Social} {Computing}}}
  \emph{(\bibinfo{series}{{CSCW} '20 {Companion}})}. \bibinfo{publisher}{ACM},
  \bibinfo{address}{New York, NY, USA}, \bibinfo{pages}{75--79}.
\newblock
\showISBNx{978-1-4503-8059-1}
\href{https://doi.org/10.1145/3406865.3418365}{doi:\nolinkurl{10.1145/3406865.3418365}}


\bibitem[Badillo-Urquiola et~al\mbox{.}(2017)]%
        {badillo-urquiola_understanding_2017}
\bibfield{author}{\bibinfo{person}{Karla~A. Badillo-Urquiola},
  \bibinfo{person}{Arup~Kumar Ghosh}, {and} \bibinfo{person}{Pamela
  Wisniewski}.} \bibinfo{year}{2017}\natexlab{}.
\newblock \showarticletitle{Understanding the {Unique} {Online} {Challenges}
  {Faced} by {Teens} in the {Foster} {Care} {System}}. In
  \bibinfo{booktitle}{\emph{Companion of the 2017 {ACM} {Conf} on {Computer}
  {Supported} {Cooperative} {Work} and {Social} {Computing}}}
  \emph{(\bibinfo{series}{{CSCW} '17 {Companion}})}. \bibinfo{publisher}{ACM},
  \bibinfo{address}{New York, NY, USA}, \bibinfo{pages}{139--142}.
\newblock
\showISBNx{978-1-4503-4688-7}
\href{https://doi.org/10.1145/3022198.3026314}{doi:\nolinkurl{10.1145/3022198.3026314}}


\bibitem[Beck(1976)]%
        {beck_cognitive_1976}
\bibfield{author}{\bibinfo{person}{Aaron~T. Beck}.}
  \bibinfo{year}{1976}\natexlab{}.
\newblock \bibinfo{booktitle}{\emph{Cognitive therapy and the emotional
  disorders}}.
\newblock \bibinfo{publisher}{Int'l Universities Press},
  \bibinfo{address}{Oxford, England}.
\newblock
\newblock
\shownote{Pages: 356}.


\bibitem[Berry and Pennebaker(2010)]%
        {berry_nonverbal_2010}
\bibfield{author}{\bibinfo{person}{Diane~S. Berry} {and}
  \bibinfo{person}{James~W. Pennebaker}.} \bibinfo{year}{2010}\natexlab{}.
\newblock \showarticletitle{Nonverbal and {Verbal} {Emotional} {Expression} and
  {Health}}.
\newblock \bibinfo{journal}{\emph{Psychotherapy and Psychosomatics}}
  \bibinfo{volume}{59}, \bibinfo{number}{1} (\bibinfo{date}{Feb.}
  \bibinfo{year}{2010}), \bibinfo{pages}{11--19}.
\newblock
\showISSN{0033-3190}
\href{https://doi.org/10.1159/000288640}{doi:\nolinkurl{10.1159/000288640}}


\bibitem[Bhattacharjee et~al\mbox{.}(2022)]%
        {bhattacharjee_i_2022-1}
\bibfield{author}{\bibinfo{person}{Ananya Bhattacharjee},
  \bibinfo{person}{Joseph~Jay Williams}, \bibinfo{person}{Karrie Chou},
  \bibinfo{person}{Justice Tomlinson}, \bibinfo{person}{Jonah Meyerhoff},
  \bibinfo{person}{Alex Mariakakis}, {and} \bibinfo{person}{Rachel Kornfield}.}
  \bibinfo{year}{2022}\natexlab{}.
\newblock \showarticletitle{"{I} {Kind} of {Bounce} off {It}":{Translating}
  {Mental} {Health} {Principles} into {Real} {Life} {Through} {Story}-{Based}
  {Text} {Messages}}.
\newblock \bibinfo{journal}{\emph{Proceedings of the ACM on Human-Computer
  Interaction}} \bibinfo{volume}{6}, \bibinfo{number}{CSCW2}
  (\bibinfo{date}{Nov.} \bibinfo{year}{2022}), \bibinfo{pages}{398:1--398:31}.
\newblock
\href{https://doi.org/10.1145/3555123}{doi:\nolinkurl{10.1145/3555123}}


\bibitem[Blaustein and Kinniburgh(2010)]%
        {blaustein_treating_2010}
\bibfield{author}{\bibinfo{person}{Margaret Blaustein} {and}
  \bibinfo{person}{Kristine~M. Kinniburgh}.} \bibinfo{year}{2010}\natexlab{}.
\newblock \bibinfo{title}{Treating traumatic stress in children and
  adolescents: how to foster resilience through attachment, self-regulation,
  and competency}.
\newblock
\showISBNx{978-1-4625-3708-2}
\urldef\tempurl%
\url{http://search.ebscohost.com/login.aspx?direct=true&scope=site&db=nlebk&db=nlabk&AN=1840304}
\showURL{%
\tempurl}


\bibitem[Blenkiron(2005)]%
        {blenkiron_stories_2005}
\bibfield{author}{\bibinfo{person}{Paul Blenkiron}.}
  \bibinfo{year}{2005}\natexlab{}.
\newblock \showarticletitle{Stories and {Analogies} in {Cognitive} {Behaviour}
  {Therapy}: {A} {Clinical} {Review}}.
\newblock \bibinfo{journal}{\emph{Behavioural and Cognitive Psychotherapy}}
  \bibinfo{volume}{33}, \bibinfo{number}{1} (\bibinfo{date}{Jan.}
  \bibinfo{year}{2005}), \bibinfo{pages}{45--59}.
\newblock
\showISSN{1352-4658, 1469-1833}
\href{https://doi.org/10.1017/S1352465804001766}{doi:\nolinkurl{10.1017/S1352465804001766}}


\bibitem[Bosgraaf et~al\mbox{.}(2020)]%
        {bosgraaf_art_2020}
\bibfield{author}{\bibinfo{person}{Liesbeth Bosgraaf}, \bibinfo{person}{Marinus
  Spreen}, \bibinfo{person}{Kim Pattiselanno}, {and} \bibinfo{person}{Susan van
  Hooren}.} \bibinfo{year}{2020}\natexlab{}.
\newblock \showarticletitle{Art Therapy for Psychosocial Problems in Children
  and Adolescents: A Systematic Narrative Review on Art Therapeutic Means and
  Forms of Expression, Therapist Behavior, and Supposed Mechanisms of Change}.
\newblock   \bibinfo{volume}{11} (\bibinfo{year}{2020}),
  \bibinfo{pages}{584685}.
\newblock
\showISSN{1664-1078}
\href{https://doi.org/10.3389/fpsyg.2020.584685}{doi:\nolinkurl{10.3389/fpsyg.2020.584685}}


\bibitem[Braun and Clarke(2006)]%
        {braun_using_2006}
\bibfield{author}{\bibinfo{person}{Virginia Braun} {and}
  \bibinfo{person}{Victoria Clarke}.} \bibinfo{year}{2006}\natexlab{}.
\newblock \showarticletitle{Using thematic analysis in psychology}.
\newblock \bibinfo{journal}{\emph{Qualitative Research in Psychology}}
  \bibinfo{volume}{3}, \bibinfo{number}{2} (\bibinfo{date}{April}
  \bibinfo{year}{2006}), \bibinfo{pages}{77--101}.
\newblock
\showISSN{14780887}
\href{https://doi.org/10.1191/1478088706qp063oa}{doi:\nolinkurl{10.1191/1478088706qp063oa}}
\newblock
\shownote{Publisher: Routledge}.


\bibitem[Chilton(2013)]%
        {chilton_art_2013}
\bibfield{author}{\bibinfo{person}{Gioia Chilton}.}
  \bibinfo{year}{2013}\natexlab{}.
\newblock \showarticletitle{Art {Therapy} and {Flow}: {A} {Review} of the
  {Literature} and {Applications}}.
\newblock \bibinfo{journal}{\emph{Art Therapy}} \bibinfo{volume}{30},
  \bibinfo{number}{2} (\bibinfo{date}{June} \bibinfo{year}{2013}),
  \bibinfo{pages}{64--70}.
\newblock
\showISSN{0742-1656}
\href{https://doi.org/10.1080/07421656.2013.787211}{doi:\nolinkurl{10.1080/07421656.2013.787211}}
\newblock
\shownote{Publisher: Routledge}.


\bibitem[Choudhury and De(2014)]%
        {choudhury_mental_2014}
\bibfield{author}{\bibinfo{person}{Munmun~De Choudhury} {and}
  \bibinfo{person}{Sushovan De}.} \bibinfo{year}{2014}\natexlab{}.
\newblock \showarticletitle{Mental {Health} {Discourse} on reddit:
  {Self}-{Disclosure}, {Social} {Support}, and {Anonymity}}.
\newblock \bibinfo{journal}{\emph{Proceedings of the Int'l AAAI Conf on Web and
  Social Media}} \bibinfo{volume}{8}, \bibinfo{number}{1} (\bibinfo{date}{May}
  \bibinfo{year}{2014}), \bibinfo{pages}{71--80}.
\newblock
\showISSN{2334-0770}
\href{https://doi.org/10.1609/icwsm.v8i1.14526}{doi:\nolinkurl{10.1609/icwsm.v8i1.14526}}
\newblock
\shownote{Number: 1}.


\bibitem[Cohen and Wills(1985)]%
        {cohen_stress_1985}
\bibfield{author}{\bibinfo{person}{Sheldon Cohen} {and}
  \bibinfo{person}{Thomas~A. Wills}.} \bibinfo{year}{1985}\natexlab{}.
\newblock \showarticletitle{Stress, social support, and the buffering
  hypothesis}.
\newblock  \bibinfo{volume}{98}, \bibinfo{number}{2} (\bibinfo{year}{1985}),
  \bibinfo{pages}{310--357}.
\newblock
\showISSN{1939-1455}
\href{https://doi.org/10.1037/0033-2909.98.2.310}{doi:\nolinkurl{10.1037/0033-2909.98.2.310}}
\newblock
\shownote{Place: {US} Publisher: American Psychological Association}.


\bibitem[Cong et~al\mbox{.}(2021)]%
        {cong_collective_2021}
\bibfield{author}{\bibinfo{person}{Nina Cong}, \bibinfo{person}{Kevin Cheng},
  \bibinfo{person}{Haoqi Zhang}, {and} \bibinfo{person}{Ryan Louie}.}
  \bibinfo{year}{2021}\natexlab{}.
\newblock \showarticletitle{Collective {Narrative}: {Scaffolding} {Community}
  {Storytelling} through {Context}-{Awareness}}. In
  \bibinfo{booktitle}{\emph{Companion {Publication} of the 2021 {Conf} on
  {Computer} {Supported} {Cooperative} {Work} and {Social} {Computing}}}
  \emph{(\bibinfo{series}{{CSCW} '21 {Companion}})}. \bibinfo{publisher}{ACM},
  \bibinfo{address}{New York, NY, USA}, \bibinfo{pages}{40--43}.
\newblock
\showISBNx{978-1-4503-8479-7}
\href{https://doi.org/10.1145/3462204.3481747}{doi:\nolinkurl{10.1145/3462204.3481747}}


\bibitem[Cornejo et~al\mbox{.}(2016)]%
        {cornejo_vulnerability_2016}
\bibfield{author}{\bibinfo{person}{Raymundo Cornejo}, \bibinfo{person}{Robin
  Brewer}, \bibinfo{person}{Caroline Edasis}, {and} \bibinfo{person}{Anne~Marie
  Piper}.} \bibinfo{year}{2016}\natexlab{}.
\newblock \showarticletitle{Vulnerability, {Sharing}, and {Privacy}:
  {Analyzing} {Art} {Therapy} for {Older} {Adults} with {Dementia}}. In
  \bibinfo{booktitle}{\emph{Proceedings of the 19th {ACM} {Conf} on
  {Computer}-{Supported} {Cooperative} {Work} \& {Social} {Computing}}}
  \emph{(\bibinfo{series}{{CSCW} '16})}. \bibinfo{publisher}{ACM},
  \bibinfo{address}{New York, NY, USA}, \bibinfo{pages}{1572--1583}.
\newblock
\showISBNx{978-1-4503-3592-8}
\href{https://doi.org/10.1145/2818048.2819960}{doi:\nolinkurl{10.1145/2818048.2819960}}


\bibitem[Costanza-Chock(2020)]%
        {costanza-chock_design_2020}
\bibfield{author}{\bibinfo{person}{Sasha Costanza-Chock}.}
  \bibinfo{year}{2020}\natexlab{}.
\newblock \bibinfo{booktitle}{\emph{Design {Justice}: {Community}-{Led}
  {Practices} to {Build} the {Worlds} {We} {Need}}}.
\newblock
\href{https://doi.org/10.7551/mitpress/12255.001.0001}{doi:\nolinkurl{10.7551/mitpress/12255.001.0001}}


\bibitem[Daily and Picard(2004)]%
        {daily_inner-active_2004}
\bibfield{author}{\bibinfo{person}{Shaundra~Bryant Daily} {and}
  \bibinfo{person}{Rosalind Picard}.} \bibinfo{year}{2004}\natexlab{}.
\newblock \showarticletitle{{{INNER-active}} Journal}. In
  \bibinfo{booktitle}{\emph{Proceedings of the 1st {{ACM}} Workshop on
  {{Story}} Representation, Mechanism and Context}}
  \emph{(\bibinfo{series}{{{SRMC}} '04})}. \bibinfo{publisher}{{ACM}},
  \bibinfo{address}{{New York, NY, USA}}, \bibinfo{pages}{51--54}.
\newblock
\showISBNx{978-1-58113-931-0}
\href{https://doi.org/10.1145/1026633.1026645}{doi:\nolinkurl{10.1145/1026633.1026645}}


\bibitem[Daily and Picard(2007)]%
        {daily_girls_nodate}
\bibfield{author}{\bibinfo{person}{Shaundra~B Daily} {and}
  \bibinfo{person}{Rosalind~W Picard}.} \bibinfo{year}{2007}\natexlab{}.
\newblock \showarticletitle{Girls {{Involved}} in {{Real Life Sharing}}:
  {{Utilizing Technology}} To}.
\newblock  (\bibinfo{year}{2007}), \bibinfo{pages}{20}.
\newblock


\bibitem[Devito et~al\mbox{.}(2019)]%
        {devito_social_2019}
\bibfield{author}{\bibinfo{person}{Michael~A. Devito},
  \bibinfo{person}{Ashley~Marie Walker}, \bibinfo{person}{Jeremy Birnholtz},
  \bibinfo{person}{Kathryn Ringland}, \bibinfo{person}{Kathryn Macapagal},
  \bibinfo{person}{Ashley Kraus}, \bibinfo{person}{Sean Munson},
  \bibinfo{person}{Calvin Liang}, {and} \bibinfo{person}{Herman Saksono}.}
  \bibinfo{year}{2019}\natexlab{}.
\newblock \showarticletitle{Social {Technologies} for {Digital} {Wellbeing}
  {Among} {Marginalized} {Communities}}. In \bibinfo{booktitle}{\emph{Companion
  {Publication} of the 2019 {Conf} on {Computer} {Supported} {Cooperative}
  {Work} and {Social} {Computing}}} \emph{(\bibinfo{series}{{CSCW} '19
  {Companion}})}. \bibinfo{publisher}{ACM}, \bibinfo{address}{New York, NY,
  USA}, \bibinfo{pages}{449--454}.
\newblock
\showISBNx{978-1-4503-6692-2}
\href{https://doi.org/10.1145/3311957.3359442}{doi:\nolinkurl{10.1145/3311957.3359442}}


\bibitem[Felitti(2002)]%
        {felitti_relation_2002}
\bibfield{author}{\bibinfo{person}{Vincent~J Felitti}.}
  \bibinfo{year}{2002}\natexlab{}.
\newblock \showarticletitle{The Relation Between Adverse Childhood Experiences
  and Adult Health: Turning Gold into Lead}.
\newblock  \bibinfo{volume}{6}, \bibinfo{number}{1} (\bibinfo{year}{2002}),
  \bibinfo{pages}{44--47}.
\newblock
\showISSN{1552-5767}
\urldef\tempurl%
\url{https://www.ncbi.nlm.nih.gov/pmc/articles/PMC6220625/}
\showURL{%
\tempurl}


\bibitem[Foa and Rauch(2004)]%
        {foa_cognitive_2004}
\bibfield{author}{\bibinfo{person}{Edna~B. Foa} {and} \bibinfo{person}{Sheila
  A.~M. Rauch}.} \bibinfo{year}{2004}\natexlab{}.
\newblock \showarticletitle{Cognitive changes during prolonged exposure versus
  prolonged exposure plus cognitive restructuring in female assault survivors
  with posttraumatic stress disorder}.
\newblock \bibinfo{journal}{\emph{Journal of Consulting and Clinical
  Psychology}} \bibinfo{volume}{72}, \bibinfo{number}{5} (\bibinfo{date}{Oct.}
  \bibinfo{year}{2004}), \bibinfo{pages}{879--884}.
\newblock
\showISSN{0022-006X}
\href{https://doi.org/10.1037/0022-006X.72.5.879}{doi:\nolinkurl{10.1037/0022-006X.72.5.879}}


\bibitem[Haeyen and Noorthoorn(2021)]%
        {haeyen_validity_2021}
\bibfield{author}{\bibinfo{person}{Suzanne Haeyen} {and} \bibinfo{person}{Eric
  Noorthoorn}.} \bibinfo{year}{2021}\natexlab{}.
\newblock \showarticletitle{Validity of the {Self}-{Expression} and {Emotion}
  {Regulation} in {Art} {Therapy} {Scale} ({SERATS})}.
\newblock \bibinfo{journal}{\emph{PLOS ONE}} \bibinfo{volume}{16},
  \bibinfo{number}{3} (\bibinfo{date}{March} \bibinfo{year}{2021}),
  \bibinfo{pages}{e0248315}.
\newblock
\showISSN{1932-6203}
\href{https://doi.org/10.1371/journal.pone.0248315}{doi:\nolinkurl{10.1371/journal.pone.0248315}}
\newblock
\shownote{Publisher: Public Library of Science}.


\bibitem[Hayward({[n.\,d.]})]%
        {hayward_what_nodate}
\bibfield{author}{\bibinfo{person}{Mark Hayward}.}
  \bibinfo{year}{[n.\,d.]}\natexlab{}.
\newblock \showarticletitle{What is narrative therapy}.
\newblock  (\bibinfo{year}{[n.\,d.]}).
\newblock


\bibitem[Herman(2015)]%
        {herman_trauma_2015}
\bibfield{author}{\bibinfo{person}{Judith~Lewis Herman}.}
  \bibinfo{year}{2015}\natexlab{}.
\newblock \bibinfo{booktitle}{\emph{Trauma and recovery: the aftermath of
  violence, from domestic abuse to political terror}}.
\newblock \bibinfo{publisher}{Basic Books, a member of the Perseus Books
  Group}.
\newblock
\showISBNx{978-0-465-06171-6 978-0-465-08730-3}
\newblock
\shownote{Section: ix, 326 pages ; 21 cm}.


\bibitem[Kaplan et~al\mbox{.}(2011)]%
        {kaplan_internet_2011}
\bibfield{author}{\bibinfo{person}{Katy Kaplan}, \bibinfo{person}{Mark~S.
  Salzer}, \bibinfo{person}{Phyllis Solomon}, \bibinfo{person}{Eugene
  Brusilovskiy}, {and} \bibinfo{person}{Pamela Cousounis}.}
  \bibinfo{year}{2011}\natexlab{}.
\newblock \showarticletitle{Internet peer support for individuals with
  psychiatric disabilities: {A} randomized controlled trial}.
\newblock \bibinfo{journal}{\emph{Social Science \& Medicine (1982)}}
  \bibinfo{volume}{72}, \bibinfo{number}{1} (\bibinfo{date}{Jan.}
  \bibinfo{year}{2011}), \bibinfo{pages}{54--62}.
\newblock
\showISSN{1873-5347}
\href{https://doi.org/10.1016/j.socscimed.2010.09.037}{doi:\nolinkurl{10.1016/j.socscimed.2010.09.037}}


\bibitem[Kawachi and Berkman(2001)]%
        {kawachi_social_2001}
\bibfield{author}{\bibinfo{person}{Ichiro Kawachi} {and}
  \bibinfo{person}{Lisa~F. Berkman}.} \bibinfo{year}{2001}\natexlab{}.
\newblock \showarticletitle{Social ties and mental health}.
\newblock   \bibinfo{volume}{78, 3} (\bibinfo{year}{2001}),
  \bibinfo{pages}{458--467}.
\newblock
\href{https://doi.org/10.1093/jurban/78.3.458}{doi:\nolinkurl{10.1093/jurban/78.3.458}}


\bibitem[Lee and Wagner(2002)]%
        {lee_effect_2002}
\bibfield{author}{\bibinfo{person}{Victoria Lee} {and} \bibinfo{person}{Hugh
  Wagner}.} \bibinfo{year}{2002}\natexlab{}.
\newblock \showarticletitle{The {Effect} of {Social} {Presence} on the {Facial}
  and {Verbal} {Expression} of {Emotion} and the {Interrelationships} {Among}
  {Emotion} {Components}}.
\newblock \bibinfo{journal}{\emph{Journal of Nonverbal Behavior}}
  \bibinfo{volume}{26}, \bibinfo{number}{1} (\bibinfo{date}{March}
  \bibinfo{year}{2002}), \bibinfo{pages}{3--25}.
\newblock
\showISSN{1573-3653}
\href{https://doi.org/10.1023/A:1014479919684}{doi:\nolinkurl{10.1023/A:1014479919684}}


\bibitem[Lenkowsky(1987)]%
        {lenkowsky_bibliotherapy_1987}
\bibfield{author}{\bibinfo{person}{Ronald~S. Lenkowsky}.}
  \bibinfo{year}{1987}\natexlab{}.
\newblock \showarticletitle{Bibliotherapy: {A} {Review} and {Analysis} of the
  {Literature}}.
\newblock \bibinfo{journal}{\emph{The Journal of Special Education}}
  \bibinfo{volume}{21}, \bibinfo{number}{2} (\bibinfo{date}{May}
  \bibinfo{year}{1987}), \bibinfo{pages}{123--132}.
\newblock
\showISSN{0022-4669}
\href{https://doi.org/10.1177/002246698702100211}{doi:\nolinkurl{10.1177/002246698702100211}}
\newblock
\shownote{Publisher: SAGE Publications Inc}.


\bibitem[Lilly(2023)]%
        {lilly_putting_2023}
\bibfield{author}{\bibinfo{person}{Jenn Lilly}.}
  \bibinfo{year}{2023}\natexlab{}.
\newblock \showarticletitle{"{Putting} my life into a story": {A} {Preliminary}
  {Evaluation} of a {Digital} {Narrative} {Intervention} {Combining}
  {Participatory} {Video} and {Narrative} {Therapy}}.
\newblock \bibinfo{journal}{\emph{Journal for Social Action in Counseling \&
  Psychology}} \bibinfo{volume}{15}, \bibinfo{number}{2}
  (\bibinfo{year}{2023}), \bibinfo{pages}{50--61}.
\newblock
\showISSN{2159-8142}
\href{https://doi.org/10.33043/JSACP.15.2.50-61}{doi:\nolinkurl{10.33043/JSACP.15.2.50-61}}
\newblock
\shownote{Number: 2}.


\bibitem[Lopes et~al\mbox{.}(2014)]%
        {lopes_narrative_2014}
\bibfield{author}{\bibinfo{person}{Rodrigo~T. Lopes},
  \bibinfo{person}{Miguel~M. Gonçalves}, \bibinfo{person}{Paulo~P.P. Machado},
  \bibinfo{person}{Dana Sinai}, \bibinfo{person}{Tiago Bento}, {and}
  \bibinfo{person}{João Salgado}.} \bibinfo{year}{2014}\natexlab{}.
\newblock \showarticletitle{Narrative {Therapy} vs. {Cognitive}-{Behavioral}
  {Therapy} for moderate depression: {Empirical} evidence from a controlled
  clinical trial}.
\newblock \bibinfo{journal}{\emph{Psychotherapy Research}}
  \bibinfo{volume}{24}, \bibinfo{number}{6} (\bibinfo{date}{Nov.}
  \bibinfo{year}{2014}), \bibinfo{pages}{662--674}.
\newblock
\showISSN{1050-3307}
\href{https://doi.org/10.1080/10503307.2013.874052}{doi:\nolinkurl{10.1080/10503307.2013.874052}}
\newblock
\shownote{Publisher: Routledge \_eprint:
  https://doi.org/10.1080/10503307.2013.874052}.


\bibitem[Madryas({[n.\,d.]})]%
        {madryas_narrative_nodate}
\bibfield{author}{\bibinfo{person}{Weronika~E Madryas}.}
  \bibinfo{year}{[n.\,d.]}\natexlab{}.
\newblock \showarticletitle{Narrative art therapy, on the basis of
  bibliotherapy, as a tool supporting narrative}.
\newblock  (\bibinfo{year}{[n.\,d.]}).
\newblock


\bibitem[Maier et~al\mbox{.}(2015)]%
        {maier_giving_2015}
\bibfield{author}{\bibinfo{person}{Christian Maier}, \bibinfo{person}{Sven
  Laumer}, \bibinfo{person}{Andreas Eckhardt}, {and} \bibinfo{person}{Tim
  Weitzel}.} \bibinfo{year}{2015}\natexlab{}.
\newblock \showarticletitle{Giving too much social support: social overload on
  social networking sites}.
\newblock \bibinfo{journal}{\emph{European Journal of Information Systems}}
  \bibinfo{volume}{24}, \bibinfo{number}{5} (\bibinfo{date}{Sept.}
  \bibinfo{year}{2015}), \bibinfo{pages}{447--464}.
\newblock
\showISSN{0960-085X}
\href{https://doi.org/10.1057/ejis.2014.3}{doi:\nolinkurl{10.1057/ejis.2014.3}}
\newblock
\shownote{Publisher: Taylor \& Francis \_eprint:
  https://doi.org/10.1057/ejis.2014.3}.


\bibitem[Malchiodi(2005)]%
        {malchiodi_expressive_2005}
\bibfield{author}{\bibinfo{person}{Cathy~A. Malchiodi}.}
  \bibinfo{year}{2005}\natexlab{}.
\newblock \bibinfo{booktitle}{\emph{Expressive therapies}}.
\newblock \bibinfo{publisher}{Guilford Press}.
\newblock
\showISBNx{978-1-59385-087-6 978-1-59385-379-2}
\urldef\tempurl%
\url{http://catdir.loc.gov/catdir/toc/ecip0421/2004019048.html}
\showURL{%
\tempurl}


\bibitem[Malchiodi(2014)]%
        {malchiodi_creative_2014}
\bibfield{author}{\bibinfo{person}{Cathy~A. Malchiodi}.}
  \bibinfo{year}{2014}\natexlab{}.
\newblock \bibinfo{booktitle}{\emph{Creative Interventions with Traumatized
  Children, Second Edition}}.
\newblock \bibinfo{publisher}{Guilford Publications}.
\newblock
\showISBNx{978-1-4625-1816-6}
\newblock
\shownote{Google-Books-{ID}: n6W1BAAAQBAJ}.


\bibitem[Marrs(1995)]%
        {marrs_meta-analysis_1995}
\bibfield{author}{\bibinfo{person}{Rick~W. Marrs}.}
  \bibinfo{year}{1995}\natexlab{}.
\newblock \showarticletitle{A meta-analysis of bibliotherapy studies}.
\newblock \bibinfo{journal}{\emph{American Journal of Community Psychology}}
  \bibinfo{volume}{23}, \bibinfo{number}{6} (\bibinfo{year}{1995}),
  \bibinfo{pages}{843--870}.
\newblock
\showISSN{1573-2770}
\href{https://doi.org/10.1007/BF02507018}{doi:\nolinkurl{10.1007/BF02507018}}
\newblock
\shownote{\_eprint:
  https://onlinelibrary.wiley.com/doi/pdf/10.1007/BF02507018}.


\bibitem[Martin et~al\mbox{.}(1992)]%
        {martin_therapists_1992}
\bibfield{author}{\bibinfo{person}{Jack Martin}, \bibinfo{person}{Anne~L.
  Cummings}, {and} \bibinfo{person}{Ernest~T. Hallberg}.}
  \bibinfo{year}{1992}\natexlab{}.
\newblock \showarticletitle{Therapists' Intentional Use of Metaphor:
  {{Memorability}}, Clinical Impact, and Possible Epistemic/Motivational
  Functions}.
\newblock \bibinfo{journal}{\emph{Journal of Consulting and Clinical
  Psychology}} \bibinfo{volume}{60}, \bibinfo{number}{1}
  (\bibinfo{year}{1992}), \bibinfo{pages}{143--145}.
\newblock
\showISSN{1939-2117}
\href{https://doi.org/10.1037/0022-006X.60.1.143}{doi:\nolinkurl{10.1037/0022-006X.60.1.143}}


\bibitem[Mateas and Sengers(2003)]%
        {mateas_narrative_nodate}
\bibfield{author}{\bibinfo{person}{Michael Mateas} {and}
  \bibinfo{person}{Phoebe Sengers}.} \bibinfo{year}{2003}\natexlab{}.
\newblock \showarticletitle{Narrative {{Intelligence}}}.
\newblock  (\bibinfo{year}{2003}), \bibinfo{pages}{10}.
\newblock


\bibitem[Mcardle and Byrt(2001)]%
        {mcardle_fiction_2001}
\bibfield{author}{\bibinfo{person}{S. Mcardle} {and} \bibinfo{person}{R.
  Byrt}.} \bibinfo{year}{2001}\natexlab{}.
\newblock \showarticletitle{Fiction, poetry and mental health: expressive and
  therapeutic uses of literature}.
\newblock \bibinfo{journal}{\emph{Journal of Psychiatric and Mental Health
  Nursing}} \bibinfo{volume}{8}, \bibinfo{number}{6} (\bibinfo{year}{2001}),
  \bibinfo{pages}{517--524}.
\newblock
\showISSN{1365-2850}
\href{https://doi.org/10.1046/j.1351-0126.2001.00428.x}{doi:\nolinkurl{10.1046/j.1351-0126.2001.00428.x}}
\newblock
\shownote{\_eprint:
  https://onlinelibrary.wiley.com/doi/pdf/10.1046/j.1351-0126.2001.00428.x}.


\bibitem[McDuff et~al\mbox{.}(2012)]%
        {mcduff_affectaura_2012}
\bibfield{author}{\bibinfo{person}{Daniel McDuff}, \bibinfo{person}{Amy
  Karlson}, \bibinfo{person}{Ashish Kapoor}, \bibinfo{person}{Asta Roseway},
  {and} \bibinfo{person}{Mary Czerwinski}.} \bibinfo{year}{2012}\natexlab{}.
\newblock \showarticletitle{{AffectAura}: an intelligent system for emotional
  memory}. In \bibinfo{booktitle}{\emph{Proceedings of the {SIGCHI} {Conf} on
  {Human} {Factors} in {Computing} {Systems}}}. \bibinfo{publisher}{ACM},
  \bibinfo{address}{Austin Texas USA}, \bibinfo{pages}{849--858}.
\newblock
\showISBNx{978-1-4503-1015-4}
\href{https://doi.org/10.1145/2207676.2208525}{doi:\nolinkurl{10.1145/2207676.2208525}}


\bibitem[McHugh et~al\mbox{.}(2018)]%
        {mchugh_when_2018}
\bibfield{author}{\bibinfo{person}{Bridget~Christine McHugh},
  \bibinfo{person}{Pamela Wisniewski}, \bibinfo{person}{Mary~Beth Rosson},
  {and} \bibinfo{person}{John~M. Carroll}.} \bibinfo{year}{2018}\natexlab{}.
\newblock \showarticletitle{When social media traumatizes teens: {The} roles of
  online risk exposure, coping, and post-traumatic stress}.
\newblock \bibinfo{journal}{\emph{Internet Research}} \bibinfo{volume}{28},
  \bibinfo{number}{5} (\bibinfo{date}{Jan.} \bibinfo{year}{2018}),
  \bibinfo{pages}{1169--1188}.
\newblock
\showISSN{1066-2243}
\href{https://doi.org/10.1108/IntR-02-2017-0077}{doi:\nolinkurl{10.1108/IntR-02-2017-0077}}
\newblock
\shownote{Publisher: Emerald Publishing Limited}.


\bibitem[Mihailidis et~al\mbox{.}(2010)]%
        {mihailidis_towards_2010}
\bibfield{author}{\bibinfo{person}{Alex Mihailidis}, \bibinfo{person}{Scott
  Blunsden}, \bibinfo{person}{Jennifer Boger}, \bibinfo{person}{Brandi
  Richards}, \bibinfo{person}{Krists Zutis}, \bibinfo{person}{Laurel Young},
  {and} \bibinfo{person}{Jesse Hoey}.} \bibinfo{year}{2010}\natexlab{}.
\newblock \showarticletitle{Towards the development of a technology for art
  therapy and dementia: definition of needs and design constraints}.
\newblock \bibinfo{journal}{\emph{Arts in Psychotherapy}} \bibinfo{volume}{37},
  \bibinfo{number}{4} (\bibinfo{year}{2010}), \bibinfo{pages}{293--300}.
\newblock
\showISSN{0197-4556}
\href{https://doi.org/10.1016/j.aip.2010.05.004}{doi:\nolinkurl{10.1016/j.aip.2010.05.004}}


\bibitem[Morris et~al\mbox{.}(2010)]%
        {morris_mobile_2010}
\bibfield{author}{\bibinfo{person}{Margaret~E. Morris}, \bibinfo{person}{Qusai
  Kathawala}, \bibinfo{person}{Todd~K. Leen}, \bibinfo{person}{Ethan~E.
  Gorenstein}, \bibinfo{person}{Farzin Guilak}, \bibinfo{person}{William
  DeLeeuw}, {and} \bibinfo{person}{Michael Labhard}.}
  \bibinfo{year}{2010}\natexlab{}.
\newblock \showarticletitle{Mobile {Therapy}: {Case} {Study} {Evaluations} of a
  {Cell} {Phone} {Application} for {Emotional} {Self}-{Awareness}}.
\newblock \bibinfo{journal}{\emph{Journal of Medical Internet Research}}
  \bibinfo{volume}{12}, \bibinfo{number}{2} (\bibinfo{date}{April}
  \bibinfo{year}{2010}), \bibinfo{pages}{e1371}.
\newblock
\href{https://doi.org/10.2196/jmir.1371}{doi:\nolinkurl{10.2196/jmir.1371}}
\newblock
\shownote{Company: Journal of Medical Internet Research Distributor: Journal of
  Medical Internet Research Institution: Journal of Medical Internet Research
  Label: Journal of Medical Internet Research Publisher: JMIR Publications
  Inc., Toronto, Canada}.


\bibitem[Morris et~al\mbox{.}(2015)]%
        {morris_efficacy_2015}
\bibfield{author}{\bibinfo{person}{Robert~R Morris}, \bibinfo{person}{Stephen~M
  Schueller}, {and} \bibinfo{person}{Rosalind~W Picard}.}
  \bibinfo{year}{2015}\natexlab{}.
\newblock \showarticletitle{Efficacy of a {Web}-{Based}, {Crowdsourced}
  {Peer}-{To}-{Peer} {Cognitive} {Reappraisal} {Platform} for {Depression}:
  {Randomized} {Controlled} {Trial}}.
\newblock \bibinfo{journal}{\emph{Journal of Medical Internet Research}}
  \bibinfo{volume}{17}, \bibinfo{number}{3} (\bibinfo{date}{March}
  \bibinfo{year}{2015}), \bibinfo{pages}{e72}.
\newblock
\showISSN{1439-4456}
\href{https://doi.org/10.2196/jmir.4167}{doi:\nolinkurl{10.2196/jmir.4167}}


\bibitem[Nurser et~al\mbox{.}(2018)]%
        {nurser_personal_2018}
\bibfield{author}{\bibinfo{person}{Kate~P. Nurser}, \bibinfo{person}{Imogen
  Rushworth}, \bibinfo{person}{Tom Shakespeare}, {and} \bibinfo{person}{Deirdre
  Williams}.} \bibinfo{year}{2018}\natexlab{}.
\newblock \showarticletitle{Personal storytelling in mental health recovery}.
\newblock \bibinfo{journal}{\emph{Mental Health Review Journal}}
  \bibinfo{volume}{23}, \bibinfo{number}{1} (\bibinfo{date}{Jan.}
  \bibinfo{year}{2018}), \bibinfo{pages}{25--36}.
\newblock
\showISSN{1361-9322}
\href{https://doi.org/10.1108/MHRJ-08-2017-0034}{doi:\nolinkurl{10.1108/MHRJ-08-2017-0034}}
\newblock
\shownote{Publisher: Emerald Publishing Limited}.


\bibitem[O'Leary et~al\mbox{.}(2017)]%
        {oleary_design_2017}
\bibfield{author}{\bibinfo{person}{Kathleen O'Leary}, \bibinfo{person}{Arpita
  Bhattacharya}, \bibinfo{person}{Sean~A. Munson}, \bibinfo{person}{Jacob~O.
  Wobbrock}, {and} \bibinfo{person}{Wanda Pratt}.}
  \bibinfo{year}{2017}\natexlab{}.
\newblock \showarticletitle{Design {Opportunities} for {Mental} {Health} {Peer}
  {Support} {Technologies}}. In \bibinfo{booktitle}{\emph{Proceedings of the
  2017 {ACM} {Conf} on {Computer} {Supported} {Cooperative} {Work} and {Social}
  {Computing}}} \emph{(\bibinfo{series}{{CSCW} '17})}.
  \bibinfo{publisher}{ACM}, \bibinfo{address}{New York, NY, USA},
  \bibinfo{pages}{1470--1484}.
\newblock
\showISBNx{978-1-4503-4335-0}
\href{https://doi.org/10.1145/2998181.2998349}{doi:\nolinkurl{10.1145/2998181.2998349}}


\bibitem[Otto(2000)]%
        {otto_stories_2000}
\bibfield{author}{\bibinfo{person}{Michael~W. Otto}.}
  \bibinfo{year}{2000}\natexlab{}.
\newblock \showarticletitle{Stories and metaphors in cognitive-behavior
  therapy}.
\newblock \bibinfo{journal}{\emph{Cognitive and Behavioral Practice}}
  \bibinfo{volume}{7}, \bibinfo{number}{2} (\bibinfo{date}{March}
  \bibinfo{year}{2000}), \bibinfo{pages}{166--172}.
\newblock
\showISSN{10777229}
\href{https://doi.org/10.1016/S1077-7229(00)80027-9}{doi:\nolinkurl{10.1016/S1077-7229(00)80027-9}}


\bibitem[Peng et~al\mbox{.}(2018)]%
        {peng_trip_2018}
\bibfield{author}{\bibinfo{person}{Fengjiao Peng},
  \bibinfo{person}{Veronica~Crista LaBelle}, \bibinfo{person}{Emily~Christen
  Yue}, {and} \bibinfo{person}{Rosalind~W. Picard}.}
  \bibinfo{year}{2018}\natexlab{}.
\newblock \showarticletitle{A {Trip} to the {Moon}: {Personalized} {Animated}
  {Movies} for {Self}-reflection}. In \bibinfo{booktitle}{\emph{Proceedings of
  the 2018 {CHI} {Conf} on {Human} {Factors} in {Computing} {Systems}}}.
  \bibinfo{publisher}{ACM}, \bibinfo{address}{New York, NY, USA},
  \bibinfo{pages}{1--10}.
\newblock
\showISBNx{978-1-4503-5620-6}
\urldef\tempurl%
\url{https://doi.org/10.1145/3173574.3173827}
\showURL{%
\tempurl}


\bibitem[Progga and Rubya(2023)]%
        {progga_just_2023}
\bibfield{author}{\bibinfo{person}{Farhat~Tasnim Progga} {and}
  \bibinfo{person}{Sabirat Rubya}.} \bibinfo{year}{2023}\natexlab{}.
\newblock \showarticletitle{"just like therapy!": {Investigating} the
  {Potential} of {Storytelling} in {Online} {Postpartum} {Depression}
  {Communities}}. In \bibinfo{booktitle}{\emph{Companion {Proceedings} of the
  2023 {ACM} {Int'l} {Conf} on {Supporting} {Group} {Work}}}
  \emph{(\bibinfo{series}{{GROUP} '23})}. \bibinfo{publisher}{ACM},
  \bibinfo{address}{New York, NY, USA}, \bibinfo{pages}{18--20}.
\newblock
\showISBNx{978-1-4503-9945-6}
\href{https://doi.org/10.1145/3565967.3570977}{doi:\nolinkurl{10.1145/3565967.3570977}}


\bibitem[Rajcic and McCormack(2020)]%
        {rajcic_mirror_2020}
\bibfield{author}{\bibinfo{person}{Nina Rajcic} {and} \bibinfo{person}{Jon
  McCormack}.} \bibinfo{year}{2020}\natexlab{}.
\newblock \showarticletitle{Mirror {Ritual}: {An} {Affective} {Interface} for
  {Emotional} {Self}-{Reflection}}. In \bibinfo{booktitle}{\emph{Proceedings of
  the 2020 {CHI} {Conf} on {Human} {Factors} in {Computing} {Systems}}}.
  \bibinfo{publisher}{ACM}, \bibinfo{address}{Honolulu HI USA},
  \bibinfo{pages}{1--13}.
\newblock
\showISBNx{978-1-4503-6708-0}
\href{https://doi.org/10.1145/3313831.3376625}{doi:\nolinkurl{10.1145/3313831.3376625}}


\bibitem[Ricks et~al\mbox{.}(2014)]%
        {ricks_my_2014}
\bibfield{author}{\bibinfo{person}{Lacey Ricks}, \bibinfo{person}{Sarah
  Kitchens}, \bibinfo{person}{Tonia Goodrich}, {and} \bibinfo{person}{Elizabeth
  Hancock}.} \bibinfo{year}{2014}\natexlab{}.
\newblock \showarticletitle{My {Story}: {The} {Use} of {Narrative} {Therapy} in
  {Individual} and {Group} {Counseling}}.
\newblock \bibinfo{journal}{\emph{Journal of Creativity in Mental Health}}
  \bibinfo{volume}{9}, \bibinfo{number}{1} (\bibinfo{date}{Jan.}
  \bibinfo{year}{2014}), \bibinfo{pages}{99--110}.
\newblock
\showISSN{15401383}
\href{https://doi.org/10.1080/15401383.2013.870947}{doi:\nolinkurl{10.1080/15401383.2013.870947}}
\newblock
\shownote{Publisher: Taylor \& Francis Ltd}.


\bibitem[Saksono et~al\mbox{.}(2021)]%
        {saksono_storymap_2021}
\bibfield{author}{\bibinfo{person}{Herman Saksono}, \bibinfo{person}{Carmen
  Castaneda-Sceppa}, \bibinfo{person}{Jessica~A. Hoffman},
  \bibinfo{person}{Magy Seif El-Nasr}, {and} \bibinfo{person}{Andrea Parker}.}
  \bibinfo{year}{2021}\natexlab{}.
\newblock \showarticletitle{{StoryMap}: {Using} {Social} {Modeling} and
  {Self}-{Modeling} to {Support} {Physical} {Activity} {Among} {Families} of
  {Low}-{SES} {Backgrounds}}. In \bibinfo{booktitle}{\emph{Proceedings of the
  2021 {CHI} {Conf} on {Human} {Factors} in {Computing} {Systems}}}
  \emph{(\bibinfo{series}{{CHI} '21})}. \bibinfo{publisher}{ACM},
  \bibinfo{address}{New York, NY, USA}, \bibinfo{pages}{1--14}.
\newblock
\showISBNx{978-1-4503-8096-6}
\href{https://doi.org/10.1145/3411764.3445087}{doi:\nolinkurl{10.1145/3411764.3445087}}


\bibitem[Saunders and Adams(2014)]%
        {saunders_epidemiology_2014}
\bibfield{author}{\bibinfo{person}{Benjamin~E. Saunders} {and}
  \bibinfo{person}{Zachary~W. Adams}.} \bibinfo{year}{2014}\natexlab{}.
\newblock \showarticletitle{Epidemiology of Traumatic Experiences in
  Childhood}.
\newblock  \bibinfo{volume}{23}, \bibinfo{number}{2} (\bibinfo{year}{2014}),
  \bibinfo{pages}{167--184}.
\newblock
\showISSN{1056-4993}
\href{https://doi.org/10.1016/j.chc.2013.12.003}{doi:\nolinkurl{10.1016/j.chc.2013.12.003}}


\bibitem[Shen et~al\mbox{.}(2023)]%
        {shen_modeling_2023}
\bibfield{author}{\bibinfo{person}{Jocelyn Shen}, \bibinfo{person}{Maarten
  Sap}, \bibinfo{person}{Pedro Colon-Hernandez}, \bibinfo{person}{Hae~Won
  Park}, {and} \bibinfo{person}{Cynthia Breazeal}.}
  \bibinfo{year}{2023}\natexlab{}.
\newblock \bibinfo{title}{Modeling {Empathic} {Similarity} in {Personal}
  {Narratives}}.
\newblock
\href{https://doi.org/10.48550/arXiv.2305.14246}{doi:\nolinkurl{10.48550/arXiv.2305.14246}}
\newblock
\shownote{arXiv:2305.14246 [cs]}.


\bibitem[Silvia(2005)]%
        {silvia_emotional_2005}
\bibfield{author}{\bibinfo{person}{Paul~J. Silvia}.}
  \bibinfo{year}{2005}\natexlab{}.
\newblock \showarticletitle{Emotional {Responses} to {Art}: {From} {Collation}
  and {Arousal} to {Cognition} and {Emotion}}.
\newblock \bibinfo{journal}{\emph{Review of General Psychology}}
  \bibinfo{volume}{9}, \bibinfo{number}{4} (\bibinfo{date}{Dec.}
  \bibinfo{year}{2005}), \bibinfo{pages}{342--357}.
\newblock
\showISSN{1089-2680}
\href{https://doi.org/10.1037/1089-2680.9.4.342}{doi:\nolinkurl{10.1037/1089-2680.9.4.342}}
\newblock
\shownote{Publisher: SAGE Publications Inc}.


\bibitem[Smith and Burkhalter(1987)]%
        {smith_use_1987}
\bibfield{author}{\bibinfo{person}{Darrell Smith} {and}
  \bibinfo{person}{Johnny~K. Burkhalter}.} \bibinfo{year}{1987}\natexlab{}.
\newblock \showarticletitle{The use of bibliotherapy in clinical practice}.
\newblock \bibinfo{journal}{\emph{Journal of Mental Health Counseling}}
  \bibinfo{volume}{9}, \bibinfo{number}{3} (\bibinfo{year}{1987}),
  \bibinfo{pages}{184--190}.
\newblock
\showISSN{2163-5749}
\newblock
\shownote{Place: US Publisher: American Mental Health Counselors Assn}.


\bibitem[Sprecher and Fehr(2005)]%
        {sprecher_compassionate_2005}
\bibfield{author}{\bibinfo{person}{Susan Sprecher} {and}
  \bibinfo{person}{Beverley Fehr}.} \bibinfo{year}{2005}\natexlab{}.
\newblock \showarticletitle{Compassionate love for close others and humanity}.
\newblock \bibinfo{journal}{\emph{Journal of Social and Personal
  Relationships}} \bibinfo{volume}{22}, \bibinfo{number}{5}
  (\bibinfo{date}{Oct.} \bibinfo{year}{2005}), \bibinfo{pages}{629--651}.
\newblock
\showISSN{0265-4075, 1460-3608}
\href{https://doi.org/10.1177/0265407505056439}{doi:\nolinkurl{10.1177/0265407505056439}}


\bibitem[Sweeney(2021)]%
        {sweeney_sandtray_2021}
\bibfield{author}{\bibinfo{person}{Daniel~S. Sweeney}.}
  \bibinfo{year}{2021}\natexlab{}.
\newblock \showarticletitle{Sandtray therapy}.
\newblock In \bibinfo{booktitle}{\emph{Play therapy with children: Modalities
  for change}}. \bibinfo{publisher}{American Psychological Association},
  \bibinfo{pages}{9--24}.
\newblock
\showISBNx{978-1-4338-3359-5 978-1-4338-3459-2}
\href{https://doi.org/10.1037/0000217-002}{doi:\nolinkurl{10.1037/0000217-002}}


\bibitem[Torre and Lieberman(2018)]%
        {torre_putting_2018}
\bibfield{author}{\bibinfo{person}{Jared~B. Torre} {and}
  \bibinfo{person}{Matthew~D. Lieberman}.} \bibinfo{year}{2018}\natexlab{}.
\newblock \showarticletitle{Putting {Feelings} {Into} {Words}: {Affect}
  {Labeling} as {Implicit} {Emotion} {Regulation}}.
\newblock \bibinfo{journal}{\emph{Emotion Review}} \bibinfo{volume}{10},
  \bibinfo{number}{2} (\bibinfo{date}{April} \bibinfo{year}{2018}),
  \bibinfo{pages}{116--124}.
\newblock
\showISSN{1754-0739}
\href{https://doi.org/10.1177/1754073917742706}{doi:\nolinkurl{10.1177/1754073917742706}}
\newblock
\shownote{Publisher: SAGE Publications}.


\bibitem[Tucci et~al\mbox{.}(2019)]%
        {tucci_handbook_2019}
\bibfield{author}{\bibinfo{person}{Joe Tucci}, \bibinfo{person}{Janise
  Mitchell}, {and} \bibinfo{person}{Edward~C. Tronick}.}
  \bibinfo{year}{2019}\natexlab{}.
\newblock \bibinfo{booktitle}{\emph{The {Handbook} of {Therapeutic} {Care} for
  {Children}: {Evidence}-{Informed} {Approaches} to {Working} with
  {Traumatized} {Children} and {Adolescents} in {Foster}, {Kinship} and
  {Adoptive} {Care}}}.
\newblock \bibinfo{publisher}{Jessica Kingsley Publishers}.
\newblock
\showISBNx{978-1-78450-554-7}
\newblock
\shownote{Google-Books-ID: 68mPDwAAQBAJ}.


\bibitem[Ulrich(2003)]%
        {ulrich_kj_2003}
\bibfield{author}{\bibinfo{person}{Karl Ulrich}.}
  \bibinfo{year}{2003}\natexlab{}.
\newblock \bibinfo{booktitle}{\emph{{KJ} {Diagrams}}}.
\newblock \bibinfo{type}{{T}echnical {R}eport}. \bibinfo{institution}{The
  Wharton School, University of Pennsylvania}, \bibinfo{address}{Philadelphia,
  PA}.
\newblock


\bibitem[Van~der Kolk(2014)]%
        {van_der_kolk_body_2014}
\bibfield{author}{\bibinfo{person}{Bessel Van~der Kolk}.}
  \bibinfo{year}{2014}\natexlab{}.
\newblock \bibinfo{booktitle}{\emph{The Body Keeps the Score}}.
\newblock \bibinfo{publisher}{Penguin Publishing Group}.
\newblock
\showISBNx{978-1-101-60830-2}
\urldef\tempurl%
\url{http://api.overdrive.com/v1/collections/v1L2BowAAAC4HAAA1k/products/d1643ff9-bf57-4025-9cf5-6c432a2f101a}
\showURL{%
\tempurl}
\newblock
\shownote{{OCLC}: 1002099363}.


\bibitem[White and Epston(1990)]%
        {white_narrative_1990}
\bibfield{author}{\bibinfo{person}{Michael White} {and} \bibinfo{person}{David
  Epston}.} \bibinfo{year}{1990}\natexlab{}.
\newblock \bibinfo{booktitle}{\emph{Narrative {Means} to {Therapeutic} {Ends}}
  (\bibinfo{edition}{1st edition} ed.)}.
\newblock \bibinfo{publisher}{W. W. Norton \& Company}, \bibinfo{address}{New
  York}.
\newblock
\showISBNx{978-0-393-70098-5}


\bibitem[Yalom(1995)]%
        {yalom_theory_1995}
\bibfield{author}{\bibinfo{person}{Irvin~D. Yalom}.}
  \bibinfo{year}{1995}\natexlab{}.
\newblock \bibinfo{booktitle}{\emph{The theory and practice of group
  psychotherapy} (\bibinfo{edition}{fourth edition} ed.)}.
\newblock \bibinfo{publisher}{Basic Books, A Division of
  {HarperCollinsPublishers}}.
\newblock
\showISBNx{978-0-465-08448-7}
\urldef\tempurl%
\url{http://catdir.loc.gov/catdir/enhancements/fy0831/94033555-b.html}
\showURL{%
\tempurl}
\newblock
\shownote{{OCLC}: 31133427}.


\bibitem[Yang et~al\mbox{.}(2019)]%
        {yang_channel_2019}
\bibfield{author}{\bibinfo{person}{Diyi Yang}, \bibinfo{person}{Zheng Yao},
  \bibinfo{person}{Joseph Seering}, {and} \bibinfo{person}{Robert Kraut}.}
  \bibinfo{year}{2019}\natexlab{}.
\newblock \showarticletitle{The {Channel} {Matters}: {Self}-disclosure,
  {Reciprocity} and {Social} {Support} in {Online} {Cancer} {Support}
  {Groups}}. In \bibinfo{booktitle}{\emph{Proceedings of the 2019 {CHI} {Conf}
  on {Human} {Factors} in {Computing} {Systems}}} \emph{(\bibinfo{series}{{CHI}
  '19})}. \bibinfo{publisher}{ACM}, \bibinfo{address}{New York, NY, USA},
  \bibinfo{pages}{1--15}.
\newblock
\showISBNx{978-1-4503-5970-2}
\href{https://doi.org/10.1145/3290605.3300261}{doi:\nolinkurl{10.1145/3290605.3300261}}


\bibitem[Yoosefi~Looyeh et~al\mbox{.}(2014)]%
        {yoosefi_looyeh_treating_2014}
\bibfield{author}{\bibinfo{person}{Majid Yoosefi~Looyeh},
  \bibinfo{person}{Khosrow Kamali}, \bibinfo{person}{Amin Ghasemi}, {and}
  \bibinfo{person}{Phuangphet Tonawanik}.} \bibinfo{year}{2014}\natexlab{}.
\newblock \showarticletitle{Treating social phobia in children through group
  narrative therapy}.
\newblock \bibinfo{journal}{\emph{The Arts in Psychotherapy}}
  \bibinfo{volume}{41}, \bibinfo{number}{1} (\bibinfo{date}{Feb.}
  \bibinfo{year}{2014}), \bibinfo{pages}{16--20}.
\newblock
\showISSN{01974556}
\href{https://doi.org/10.1016/j.aip.2013.11.005}{doi:\nolinkurl{10.1016/j.aip.2013.11.005}}


\bibitem[Zhang et~al\mbox{.}(2022)]%
        {zhang_separate_2022}
\bibfield{author}{\bibinfo{person}{Ben~Zefeng Zhang}, \bibinfo{person}{Tianxiao
  Liu}, \bibinfo{person}{Shanley Corvite}, \bibinfo{person}{Nazanin Andalibi},
  {and} \bibinfo{person}{Oliver~L. Haimson}.} \bibinfo{year}{2022}\natexlab{}.
\newblock \showarticletitle{Separate {Online} {Networks} {During} {Life}
  {Transitions}: {Support}, {Identity}, and {Challenges} in {Social} {Media}
  and {Online} {Communities}}.
\newblock \bibinfo{journal}{\emph{Proceedings of the ACM on Human-Computer
  Interaction}} \bibinfo{volume}{6}, \bibinfo{number}{CSCW2}
  (\bibinfo{date}{Nov.} \bibinfo{year}{2022}), \bibinfo{pages}{1--30}.
\newblock
\showISSN{2573-0142}
\href{https://doi.org/10.1145/3555559}{doi:\nolinkurl{10.1145/3555559}}


\end{thebibliography}

\appendix

\end{document}